\documentclass[twocolumn,iop,revtex4,usenatbib]{openjournal}
\usepackage[T1]{fontenc}

\DeclareRobustCommand{\VAN}[3]{#2}
\let\VANthebibliography\thebibliography
\def\thebibliography{\DeclareRobustCommand{\VAN}[3]{##3}\VANthebibliography}

\usepackage[bitstream-charter]{mathdesign}
\usepackage{graphicx}	
\usepackage{soul}
\usepackage{microtype}

\usepackage[title]{appendix}
\usepackage{hyperref}	
\hypersetup{colorlinks=true,linkcolor=blue,citecolor=blue,filecolor=blue,urlcolor=blue}

\newcommand{\ltsima}{$\; \buildrel < \over \sim \;$}
\newcommand{\lsim}{\lower.5ex\hbox{\ltsima}}
\newcommand{\gtsima}{$\; \buildrel > \over \sim \;$}
\newcommand{\gsim}{\lower.5ex\hbox{\gtsima}}
\newcommand{\bra}{\langle}
\newcommand{\ket}{\rangle}

\newcommand{\dd}{\mathrm{d}}

\newcommand{\ci}{\mathrm{i}}

\DeclareSymbolFont{usualmathcal}{OMS}{cmsy}{m}{n}
\DeclareSymbolFontAlphabet{\mathcal}{usualmathcal}

\begin{document}

\title{A unified linear intrinsic alignment model for elliptical and disc galaxies and the resulting ellipticity spectra}

\author{Basundhara Ghosh$^{1,\star}$}
\author{Kai Nu{\ss}baumer$^{2}$}
\author{Eileen Sophie Giesel$^{2}$}
\author{Bj{\"o}rn Malte Sch{\"a}fer$^{2,\sharp}$}
\thanks{$^\star$ \href{mailto:basundharag@iisc.ac.in}{basundharag@iisc.ac.in}}
\thanks{$^\sharp$ \href{mailto:bjoern.malte.schaefer@uni-heidelberg.de}{bjoern.malte.schaefer@uni-heidelberg.de}}

\affiliation{
$^{1}$ Department of Physics, Indian Institute of Science, Bengaluru 560012, Karnataka, India
\\
$^{2}$Zentrum f{\"u}r Astronomie der Universit{\"a}t Heidelberg, Astronomisches Rechen-Institut, Philosophenweg 12, 69120 Heidelberg, Germany
}

\begin{abstract}
Alignments of spiral galaxies were thought to result from tidal torquing, where tidal field of the cosmic large-scale structure exert torquing moments onto dark matter haloes, determining their angular momentum and ultimately the orientation of galactic discs. In this model, resulting intrinsic ellipticity correlations are typically present on small scales, but neither observations nor simulations have found empirical evidence; instead, simulations point at the possibility that alignments of disc galaxies follow a similar alignment model as elliptical galaxies, but with a weaker alignment amplitude. In our article we make the case for the theory of linear alignments resulting from tidal distortions of the galactic disc, investigate the physical properties of this model and derive the resulting angular ellipticity spectra, as they would appear as a contribution to weak gravitational lensing in surveys such as Euclid's. We discuss in detail on the statistical and physical properties of tidally induced alignments in disc galaxies, as they are relevant for mitigation of alignment contamination in weak lensing data, comment on the consistency between the alignment amplitude in spiral and elliptical galaxies and finally, estimate their observability with an Euclid-like survey.
\end{abstract}


\section{Introduction}
The subject of galaxy alignments is of prime importance for the interpretation of weak lensing data. In general, any dark matter halo that contains a galaxy undergoes tidal interactions due to gravitational tidal fields when embedded in the large-scale structure, and the stellar, luminous components becomes a proxy of this interaction, for instance by physical distortion of the halo away from its equilibrium shape, or by orientation, for instance relative to the halo's angular momentum. These two alignment modes are discussed in the literature and have been conceived to describe alignments in elliptical \citep{Hirata:2004gc, Hirata:2007np,Bridle:2007ft} and spiral galaxies \citep{Crittenden:2000wi, Schaefer:2011gt,Chisari:2015qga,Rodriguez-Gomez:2016jue}, respectively \citep{Schaefer:2008xd, Joachimi:2015mma, Kiessling:2015sma, Kirk:2015nma, Troxel:2014dba, Lamman:2023hsj}. Angular momentum-based alignments should generate rather strong ellipticity correlations and be easily measurable by Euclid-like \citep{Amendola:2016saw} surveys \citep{Schafer:2015qfx}. Combining the alignment models for the two major galaxy types would lead to a rich phenomenology and the question of separating the two mechanisms \citep{Tugendhat:2017qao, Reischke:2019quo}.

Tidal shearing for ellipticals and tidal torquing for spirals have been recently tested in \citet{Zjupa:2020kcg,Delgado:2023kwx} using morphologically selected elliptical and spiral galaxies from the IllustrisTNG simulation \citep{Nelson:2018uso} in the range of redshifts $0 \leq z < 1$. In this work it was found that elliptical galaxies show strong alignment signals further increasing with mass and redshift. Contrary to that, spiral galaxies exhibit a strong alignment only at high redshifts near $z=1$, and their alignment follows rather the tidal shearing instead of the tidal torquing model. Weak lensing surveys take a similar attitude, reporting a detection of a ellipticity cross-correlation between lensing and the shape of elliptical galaxies as a signature of tidal shearing. They place bounds on the cross-correlation of lensing and the shapes of spiral galaxies, implying that their alignment must be much weaker \citep{KiDS450-2017, Johnston:2018nfi,DES:2022vuu}, casting doubts on the validity of the quadratic alignment model.\par
This is at par with the \textsc{eagle} simulation \citep{Schaye:2014tpa,Crain:2015poa,Velliscig:2015ixa} and the Illustris simulation \citep{Vogelsberger:2014dza,Vogelsberger:2014kha,Hilbert:2016ylf}, whose results suggest that elliptical galaxies have a stronger alignment to that spiral galaxies of the same mass, which is also true for the host haloes of these galaxies. The MillenniumTNG simulation also finds that the alignment of spiral galaxies under a quadratic model is much weaker than that of elliptical galaxies under a linear model \citep{Delgado:2023kwx}. However, results from the MassiveBlack-II simulation \citep{Tenneti:2014bca} had shown that there is no significant difference between the direction of alignments in elliptical and spiral galaxies. Furthermore, \cite{Chisari:2015qga} state, with respect to the \textsc{horizon}-AGN simulations, that the tangential orientation of disc galaxies around spheroidals are not safe to ignore in future weak lensing surveys. Some other simulations, like \textsc{simba} \citep{Dave:2019yyq}, emphasise that alignments of galaxies with respect to host filaments are largely dictated by their formation history and stellar mass function evolution. These wide range of results suggest that the discourse around galaxy alignments, especially the comparison between ellipticals and spirals, should be thoroughly pursued.

In this work, we propose a unified alignment model that can cater to both elliptical and spiral galaxies: If such a model naturally accounts for the much weaker alignments of spirals, it would not be in conflict with observations and could account for the results from simulations. It is to be noted that we do not claim our work to be the first instance where such an alignment is mentioned, and our work is mainly intended to outline a feasible microscopic mechanism for a unified alignment mechanism for both major galaxy types. In fact, previous works on perturbative models of galaxy alignments \citep{Blazek:2017wbz,Vlah:2019byq,Chen:2023yyb} have already stated that both linear and higher-order terms exist simultaneously, and the alignment is predominantly linear or quadratic depending on the magnitude of the relative contribution from these terms. In itself, these effective field theories make the point that the dichotomy between tidal shearing and tidal torquing is too extreme. They assume that alignment models should contain the freedom to interpolate between these two bounding scenarios and state that a straightforward asssignment of alignment mechanisms to galaxy types is overly simplifying. In fact, \citet{Schaefer:2011gt} discusses the case where the relative orientation between the tidal gravitational field and the moment of inertia of the galaxy determines the relative magnitudes of tidal shearing and tidal torquing within a halo. 

Motivated by these results, we investigate the possibility of a unified linear alignment model for disc galaxies, that follows an approach analogous to our computations for the ellipticity of elliptical galaxies in \citet{Ghosh:2020zfa} and aim to establish a fundamental understanding of the respective alignment parameters. In the spirit of effective field theories, such a unified alignment model would imply that there is no need for separate alignment parameters for the two major galaxy types. Instead they would follow consistently from the internal dynamics of the galaxies and the characteristics of the stellar component. The underlying principle is that galactic discs subjected to a gravitational tidal field undergo a distortion effect that can cause them to assume an elliptical shape, resulting from the change of the otherwise circular orbit of a star in the galactic disc into an elliptical orbit. The ellipticity of the stellar orbits, which are bound by the Navarro-Frenk-White (NFW) halo potential \citep{Navarro:1995iw} should be linear in the external tidal shear.\par
An important ingredient for computing the ellipticity of galaxies is the alignment parameter which depends on the scale at which alignment due to tidal shear takes place. In case of elliptical galaxies, this scale can be assumed to be that of the S\'ersic radius, following a de Vaucouleur's profile with S\'ersic index $n=4$, whereas for spiral galaxies, it should be typically the scale radius of a galactic disc with typically exponential profile. We perform a perturbative calculation to compute the effect of a quadrupolar tidal gravitational field on a circular orbit, which naturally generates an ellipticity proportional to the magnitude of the tidal field. We then simulate the orbit of a star in an NFW-halo, which is tidally perturbed again by a quadrupolar field, and this in turn is in agreement with our perturbative results. We eventually see that the strength of alignment for disc galaxies is smaller than that of elliptical galaxies. In following this idea, we want to provide a physical idea behind linear alignments for disc galaxies as observed in simulations \citep{Hilbert:2016ylf, Jagvaral:2022zto} and overpredicted amplitude of tidal torquing models which is in disagreement with observations \citep{Heymans:2013fya,Kilbinger:2008gk}, as well as bridge to effective alignment models \citep{Blazek:2017wbz,Blazek:2011xq,Blazek:2015lfa}. With a unified model for alignments of disc and elliptical galaxies, there would be no need to differentiate between the two types, leading to simplified effective models.

Typically, we present results for a spatially flat standard $\Lambda$CDM-cosmology with conservative parameter choices: $\Omega_m = 0.3$ (with a baryonic contribution of $\Omega_b = 0.04$), $\Omega_\Lambda=1-\Omega_m$, $n_s = 0.96$, $\sigma_8 = 0.8$ and a value of $h = 0.7$. We assume an implicit summation over repeated indices. After introducing the linear alignment model for disc galaxies, investigating its physical properties and juxtaposing it with an analogous alignment model for elliptical galaxies in Sect.~\ref{sect_alignments}, we compute in Sect.~\ref{sect_ellipticity} the resulting angular ellipticity spectra for an Euclid-like survey with a redshift distribution $n(z)\dd z \propto z^2\exp(-z/z_0)$ with a median redshift of 0.9. A summary and discussion in Sect.~\ref{sect_summary} concludes our investigation.

\section{Linear alignments for disc galaxies}\label{sect_alignments}

\subsection{Scaling arguments}
Tidal fields introduce a distortion of an elliptical galaxy by differential gravitational acceleration. Assuming that the galaxy is in virial equilibrium with a velocity dispersion $\sigma^2$, the ellipticity is given by the second derivatives of the gravitational potential in units of the velocity dispersion,
\begin{equation}
\epsilon_\mathrm{el} = r_\mathrm{vir}^2\partial^2\frac{\Phi}{\sigma_\mathrm{vir}^2}.
\label{eqn_ellipticity_elliptical}
\end{equation}
As the second moments of the brightness distributions are computed out to distances of $r_\mathrm{vir}$, the constant of proportionality, i.e. the alignment parameter $D_\mathrm{IA}$, has units of area and should be related to the squared size of the galaxy. One can compute dimensionless correction factors to this relation for brightness profiles of the S{\'e}rsic-family, resulting in small correction for exponential profiles, but rather large influences in the case of de Vaucouleurs-profiles \citep{Ghosh:2020zfa,Giesel:2021ikc}. There should be a similar linear effect of tidal fields on the shape of a disc galaxy: The galactic disc extends typically to the NFW-scale radius $r_s$, which is a fraction $1/c(M_\mathrm{vir},z)$ of the virial radius $r_\mathrm{vir}$, giving rise to the phenomenon of flat rotation curves. The quantity $c(M_\mathrm{vir},z)$ is the concentration parameter which relates the virial mass $M_\mathrm{vir}$ of a halo to its redshift $z$. A quadrupolar tidal field distorts the edge of the galactic disc, compressing it in one direction and elongating it in the perpendicular direction, making it elliptical. The same scaling argument as in eqn.~(\ref{eqn_ellipticity_elliptical}) suggests a relation of the form
\begin{equation}
\epsilon_\mathrm{sp} = r_s^2\partial^2\frac{\Phi}{\upsilon_\mathrm{circ}^2}
\end{equation}
where $r_s$ denotes the edge of the galactic disc. $\upsilon_{\mathrm{circ}}$ as the circular velocity replaces $\sigma$ as the velocity dispersion, describing elasticity or susceptibility of the stellar component to tidal forces. To complete the scaling argument, we would like to state that gravitational lensing shear introduces an ellipticity
\begin{equation}
\gamma \propto 2\partial^2\frac{\Phi}{c^2}.
\end{equation}
The fact that tidal gravitational fields give rise to both lensing and intrinsic alignments causes serious problems in the analysis of lensing data. Whereas the stars as test particles of the tidal field determine the magnitude of the observable effect by either $\sigma^2$ or $\upsilon_\mathrm{circ}^2$ in intrinsic alignments, the velocity scale for photons as test particles is of course $c^2$.

\subsection{Perturbative calculation}
From scaling arguments alone one would expect an effect of tidal gravitational fields on the shape of an otherwise circular galactic disc. Whether the effect is real and present at linear order, can be shown in a straightforward perturbative calculation: The circular motion at the edge of the galactic disk is a superposition of two harmonic oscillations in perpendicular directions with the same angular frequency $\omega = \upsilon_\mathrm{circ}/r_s$ and a phase shift of $\pi/2$. The effective potential for motion in the $x,y$-plane consists of a harmonic potential responsible for the circular motion, perturbed by a quadrupole with fixed amplitude:
\begin{equation}
\Phi = \frac{\omega^2}{2}(x^2+y^2) + \frac{\Omega^2}{2}(x^2-y^2)
\label{eqn_effective_potential}
\end{equation}
where the amplitude of the quadrupolar tidal field is controlled by $\Omega\ll\omega$. Effectively, this relation encapsulates a single component of the general tidal perturbation $\partial_i\partial_j\Phi$ perturbing the potential $\Phi$
\begin{equation}
\Phi \rightarrow \Phi + \frac{1}{2}r_ir_j\:\partial_i\partial_j\Phi
\end{equation}
where we decompose the tensors $r_ir_j$ and $\partial_i\partial_j\Phi$ as
\begin{equation}
r_ir_j = (x^2+y^2)\sigma^{(0)}_{ij} + (x^2-y^2)\sigma^{(1)}_{ij} + 2xy\sigma^{(3)}_{ij}
\label{eqn_expansion_rirj}
\end{equation}
as well as
\begin{equation}
\partial_i\partial_j\Phi = \Delta\Phi\sigma^{(0)}_{ij} + (\partial^2_x-\partial^2_y)\Phi\sigma^{(1)}_{ij} + 2\partial_x\partial_j\Phi\sigma^{(3)}_{ij}
\label{eqn_expansion_phiij}
\end{equation}
with the real-valued Pauli-matrices $\sigma^{(n)}$, such that the contraction $r_ir_j\partial_i\partial_j\Phi$ becomes
\begin{equation}
r_ir_j\partial_i\partial_j\Phi = (x^2+y^2)\Delta\Phi + (x^2-y^2)(\partial^2_x-\partial^2_y)\Phi + 2xy\partial_x\partial_y\Phi
\end{equation}
We focus on the second term in eqn.~(\ref{eqn_effective_potential}) with the tidal field amplitude $\Omega^2 = (\partial^2_x-\partial^2_y)\Phi$. Interestingly, both the unperturbed circular motion and the fixed tidal perturbation are mapped onto harmonic potentials, albeit anisotropic in the second case. Considering a single oscillation,
\begin{equation}
x(t) = r_s\sin(\omega t)
\end{equation}
along the $x$-direction one obtains for the perturbed amplitude $\Delta(t)$ the equation of motion
\begin{equation}
\ddot{\Delta}(t) = -\frac{\dd\Phi}{\dd x}
\quad\mathrm{with}\quad
\frac{\dd\Phi}{\dd x} = (\omega^2+\Omega^2)x
\end{equation}
which is evaluated in the integration along the unperturbed trajectory $x(t)$, effectively in a Born-type approximation:
\begin{equation}
\ddot{\Delta}(t) = -(\omega^2+\Omega^2)r_s\sin(\omega t)
\end{equation}
Integrating twice yields a perturbative solution to the trajectory. In particular, evaluating $\Delta(t)$ at values 0 and $\pi/2$ for the phase $\omega t$ corresponds to distortions $\propto 1\pm\Omega^2/\omega^2$ of the semi-major and semi-minor axes, resulting in the ellipticity
\begin{equation}
\epsilon =
\frac{(1+\Omega^2/\omega^2)-(1-\Omega^2/\omega^2)}{(1+\Omega^2/\omega^2)+(1-\Omega^2/\omega^2)} = \frac{\Omega^2}{\omega^2} =
\frac{r_s^2}{\upsilon_\mathrm{circ}^2}\partial^2\Phi
\end{equation}
as expected from scaling arguments.

An approximate energy consideration leads to a qualitatively identical result: As the potential is perturbed by the tidal field according to the replacement
\begin{equation}
\Phi = \frac{\omega^2}{2}x^2
\quad\rightarrow\quad
\Phi = \frac{\omega^2\pm\Omega^2}{2}x^2
\end{equation}
the amplitudes in $x$- and $y$-direction that are reachable with the potential energy $V = \omega^2r_s^2/2$ are changed to
\begin{equation}
\Delta_\pm =
r_s\sqrt{\frac{\omega^2}{\omega^2+\Omega^2}} \simeq
r_s\left(1\mp\frac{\Omega^2}{2\omega^2}\right)
\end{equation}
resulting again in an ellipticity $\epsilon$
\begin{equation}
\epsilon =
\frac{\Omega^2}{2\omega^2} =
\frac{1}{2}\frac{r_s^2}{\upsilon_\mathrm{circ}^2}\partial^2\Phi
\end{equation}
The factor 2 appearing in comparison to the above argument can be traced back to the virial theorem.

Yet another interpretation of the process would be to associate a time scale to the potential of the halo, either the free-fall time scale $t_\mathrm{el} = 2\pi/\omega \simeq r_\mathrm{vir}/\sigma$ in the case of elliptical galaxies, and the orbital time scale $t_\mathrm{sp} = 2\pi/\omega \simeq r_s/\upsilon_\mathrm{circ}$ for disc galaxies, and set this in relation with the time scale that can be attributed to the (quadrupolar) tidal field, $t_\mathrm{tid} = 2\pi/\Omega$, as it is a superposition of two harmonic potentials, with $\Omega^2$ reflecting $\partial^2\Phi$. Then, the ellipticities are given by
\begin{equation}
\epsilon_\mathrm{el} = \frac{t_\mathrm{el}^2}{t_\mathrm{tid}^2}
\quad\mathrm{and}\quad
\epsilon_\mathrm{sp} = \frac{t_\mathrm{sp}^2}{t_\mathrm{tid}^2}
\end{equation}
implying that the relative alignment strength in the same tidal field should be $\epsilon_\mathrm{sp} = t_\mathrm{sp}^2/t_\mathrm{el}^2\times\epsilon_\mathrm{el}$: The ratio between free-fall time scale and orbital time scale should determine the relative magnitude of the alignments. An interesting earlier work in this context is \cite{Camelio:2015gda}, where the authors ascertain that an instantaneous external tidal field is not sufficient to explain the magnitude of the observed intrinsic alignment signals, and hence galaxy merger histories are also possible contributors to the signal. A discussion about galaxy evolution processes and their effect on our model is beyond the scope of our work, but will be an intriguing area to explore.

\subsection{Direct numerical simulation}
The perturbative calculation neglects that a perturbed orbit experiences changing gradients of the NFW-potential of the halo, and that the orbits can not be closed as Bertrand's theorem is violated. In particular the first effect is potentially interesting: Density profiles $\propto 1/r^2$ naturally give rise to flat rotation curves, as equating the centrifugal acceleration and the potential gradient according to $\upsilon_\mathrm{circ}^2/r = \dd\Phi/\dd r$ suggests constant $\upsilon_\mathrm{circ}$. While this is approximately true for NFW-profiles around $r_s$, the calculation shows decreasing velocities in radial direction. An intuitive interpretation of the violation of Bertrand's theorem in the case of a perturbed potential is that the effective frequencies $\sqrt{\omega^2\pm\Omega^2}$ in the two directions are not identical giving rise to a Lissajous figure. But for weak tidal fields, $\Omega^2\ll\omega^2$, the effect is arbitrarily small.

A direct simulation of the orbit of stars as test particles at the edge of a galactic disc with radius $r_s$ in a NFW-potential that is perturbed by a quadrupolar tidal field confirms that there is a linear reaction of the shape of the galactic disc and yields numerical results that are well in agreement with the scaling and perturbative arguments. Using dimensionless variables $x = r/r_s$ and $\tau = \omega t$ gives an equation of motion
\begin{equation}
\frac{\dd^2 x}{\dd\tau^2} = -\frac{\dd}{\dd x}\frac{\Phi}{\upsilon_\mathrm{circ}^2}
\quad\text{with}\quad
\upsilon_\mathrm{circ}^2 = (\omega r_s)^2.
\end{equation}
Specifically, an NFW-potential would have potential gradients
\begin{equation}
\frac{\dd\Phi}{\dd x} = \frac{GM_\mathrm{vir}}{Ar_s}\frac{x/(1+x) - \ln(1+x)}{x^2}
\end{equation}
which determines the circular velocity at $x=1$:
\begin{equation}
\omega^2 = \frac{GM_\mathrm{vir}}{r_\mathrm{vir}^3}\frac{B}{A}
\end{equation}
and therefore the amplitude of the potential, abbreviating $$B = 1/2-\ln2$$ and $$A = \ln(1+c(M_\mathrm{vir},z)) - \frac{c(M_\mathrm{vir},z)}{(1+c(M_\mathrm{vir},z))}$$
Summarising, the equation of motion is given explicitly by
\begin{equation}
\frac{\dd^2x}{\dd\tau^2} = -\frac{1}{B}\frac{x/(1+x) - \ln(1+x)}{x^2},
\end{equation}
which we solve numerically for a circular orbit at $x=1$ and $\upsilon_\mathrm{circ}=1$, perturbed by tidal shear fields $\Omega^2/2\:(x^2-y^2)$ and $\Omega x y$.

\begin{figure}
\centering
\includegraphics[scale=0.45]{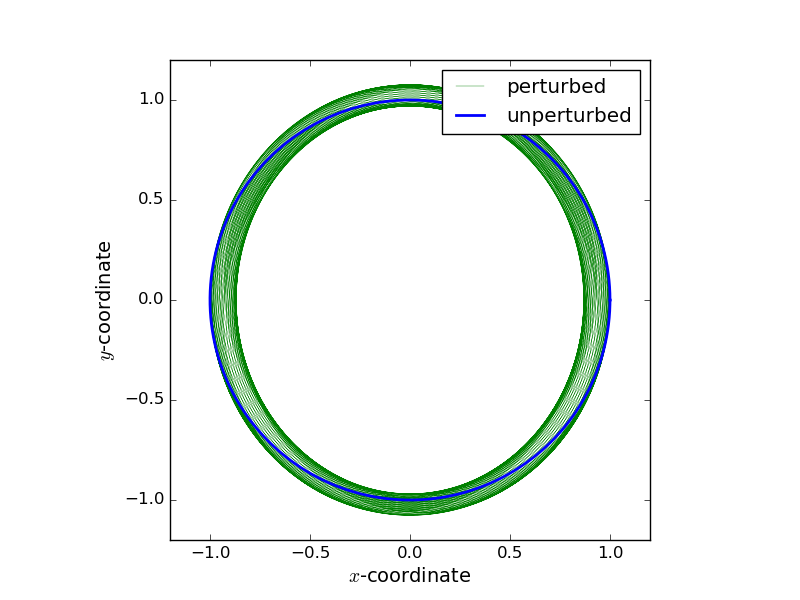}
\includegraphics[scale=0.45]{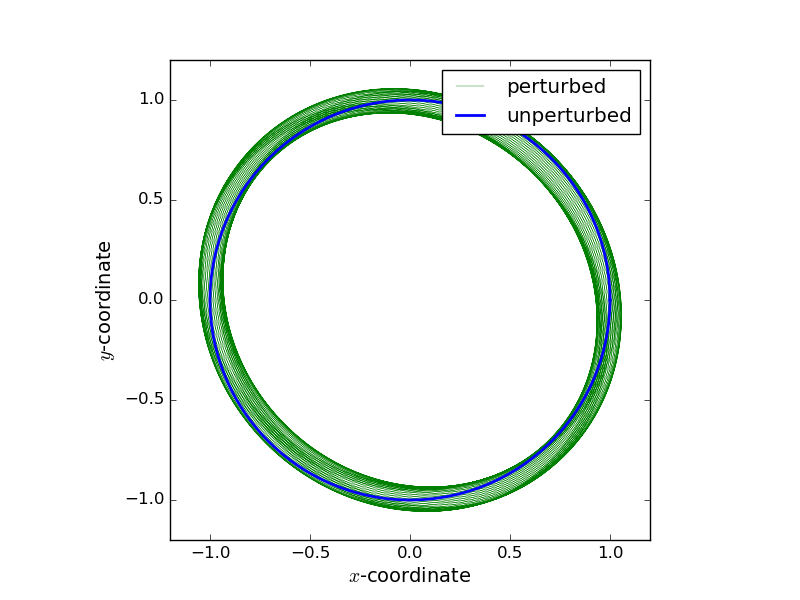}
\caption{Perturbed orbits (green) in comparison to unperturbed circular orbits (blue) at $r = r_s$ (or $x = 1$) in the potential of an NFW-halo with a tidal field $\Omega^2(x^2-y^2)/2$ leading to $\epsilon_+$ (top panel), and a tidal field $\Omega^2xy$ leading to $\epsilon_\times$ (bottom panel). The magnitude of the perturbing field is $\Omega^2 = 0.04$ in dimensionless units.}
\label{fig_perturbed}
\end{figure}

Fig.~\ref{fig_perturbed} shows the effect of tidal fields onto a circular orbit at $r_s$, specifically, the distortion into an ellipse caused by an anisotropic tidal shear field. An estimate of the ellipticity  $\epsilon$ from the semi-major and semi-minor axes is exactly the tidal shear field, in our dimensionless choice of coordinates: $\epsilon \simeq \Omega^2/\omega^2$ with $r_s = 1$ and $\upsilon_\mathrm{circ} = 1$ such that $\omega = 1/(2\pi)$. In fact, Fig.~\ref{fig_perturbed} suggests an ellipticity of $\simeq 0.1$ for a tidal shear field $\Omega^2 = 0.04$, consistent with the expectation. Isotropic tidal shear fields would cause an effect on the size of the galactic disc, which we comment on in the Appendix.

\subsection{Alignment parameters: numerical values and scaling}

\subsubsection{Perturbed dynamics in disc and elliptical galaxies}
\begin{figure}
\centering
\includegraphics[scale=0.4]{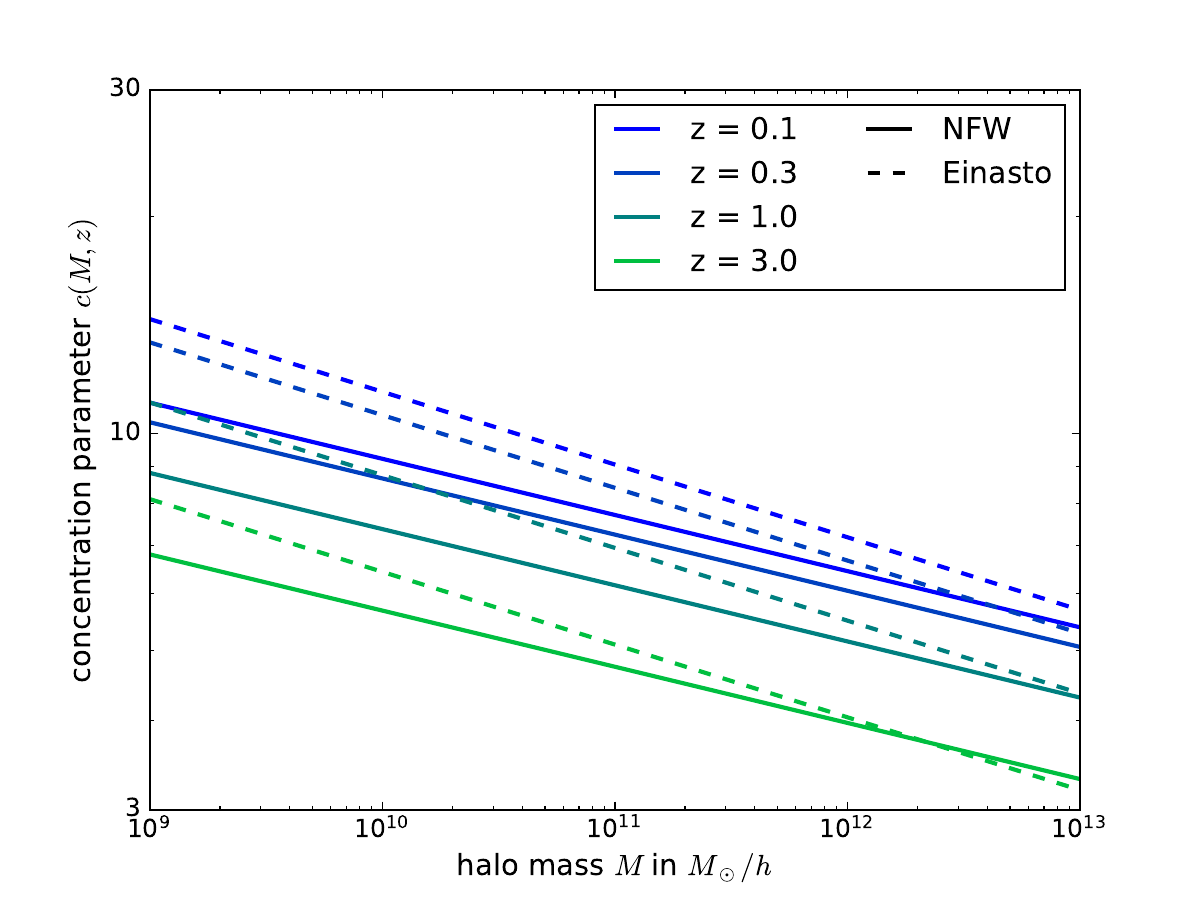}
\includegraphics[scale=0.4]{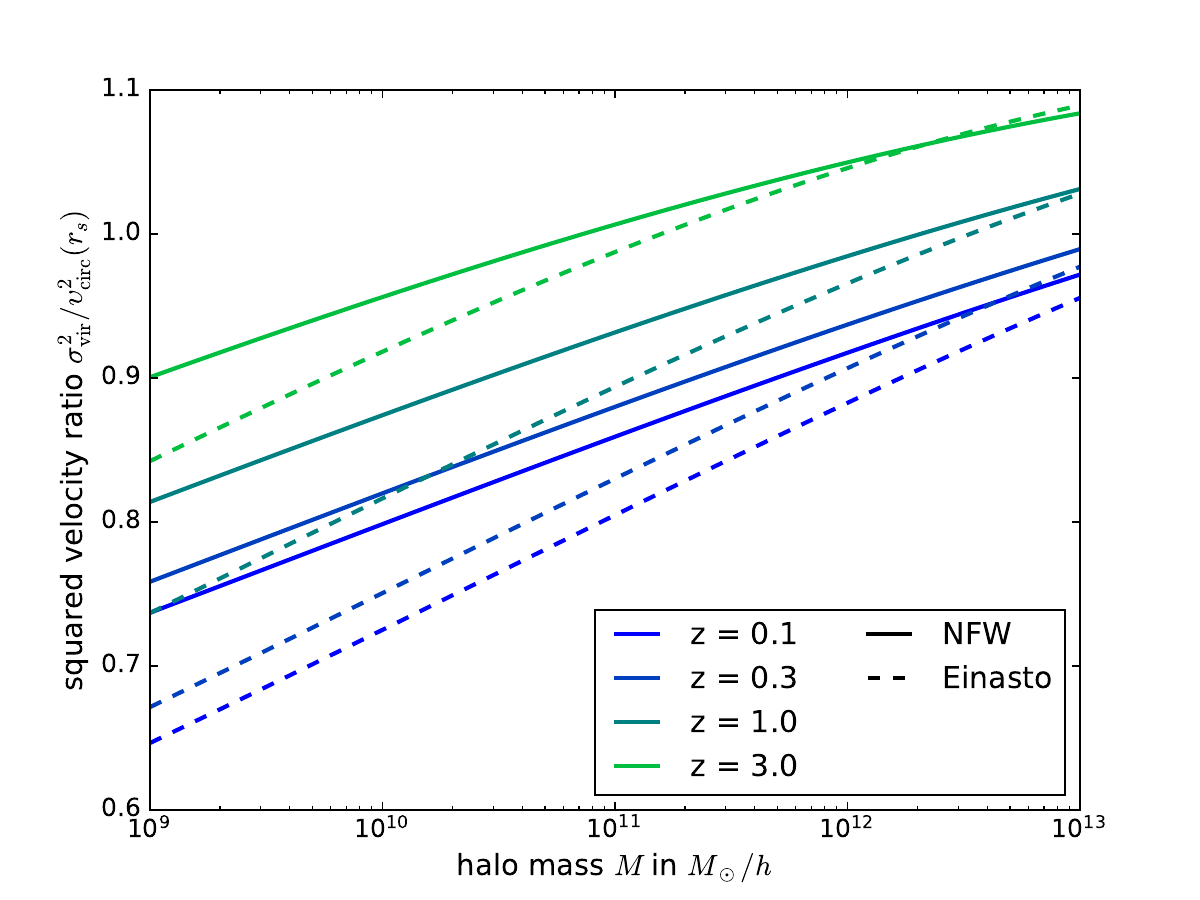}
\caption{Scaling of the concentration parameter $c(M_\mathrm{vir},z)$ with virial halo mass $M_\mathrm{vir}$ and redshift $z$ for two common halo profile families, NFW \citep{Navarro:1995iw} and Einasto \citep{1965TrAlm...5...87E} (top panel) and dependence of the squared velocity ratio $\sigma^2/\upsilon_\mathrm{circ}^2$ (bottom panel), both derived with the fitting formula from \citet{Child:2018skq}.}
\label{fig_concentration}
\end{figure}

In essence, one would expect an alignment of disc galaxies following a linear alignment model and the relation between the ellipticities of a disc galaxy and an elliptical galaxy in the same tidal gravitational field to be
\begin{equation}
\epsilon_\mathrm{sp} =
\frac{r_s^2}{r_\mathrm{vir}^2}\frac{\sigma_\mathrm{vir}^2}{\upsilon_\mathrm{circ}^2}\times\epsilon_\mathrm{el}.
\label{eqn_fundamental_scaling}
\end{equation}
Naively, the much smaller size $r_s$ of the galactic disc in comparison to the virial radius $r_\mathrm{vir}$ would be responsible for a much weaker alignment of disc galaxies relative to ellipticals. The alignment parameter for discs should be smaller by a factor of $1/c(M_\mathrm{vir},z)^2$ relative to the alignment of ellipticals, with $c(M_\mathrm{vir},z)$ being the NFW-concentration parameter. In fact, $r_s^2/r_\mathrm{vir}^2 = 1/c(M_\mathrm{vir},z)^2 \simeq 0.02\ldots0.06$, at redshifts around unity, where Euclid's galaxy distribution is peaking, and for typical masses $10^{11\ldots12}M_\odot/h$. The scaling of $c(M_\mathrm{vir},z)$ with redshift and halo mass follows the numerical model by \citet{Child:2018skq}, who condensed results from numerical simulations into phenomenological relations summarised by Fig.~\ref{fig_concentration}. On the other hand, the squared velocity ratio in the second term typically assumes values close to one, specifically $\sigma_\mathrm{vir}^2/\upsilon_\mathrm{circ}^2 \simeq 0.9\ldots0.95$ at a redshift of $z=1$ and for galaxy masses of $M = 10^{11\ldots12}M_\odot/h$ without a strong variability. Fig.~\ref{fig_concentration} shows as well how the ratio between circular velocity $\upsilon_\mathrm{circ}$ at the scale radius $r_s$ and the velocity dispersion $\sigma_\mathrm{vir}^2$
\begin{equation}
\frac{\upsilon_\mathrm{circ}^2}{\sigma_\mathrm{vir}^2} = \frac{c(M_\mathrm{vir},z)(\ln 2-1/2)}{\ln(c(M_\mathrm{vir},z)+1) - c(M_\mathrm{vir},z)/(c(M_\mathrm{vir},z)+1)},
\end{equation}
or more compactly, $\upsilon_\mathrm{circ}^2 / \sigma_\mathrm{vir}^2 = c(M_\mathrm{vir},z) B/A$. The ratio changes as a function of halo mass $M$ and redshift $z$ through the concentration parameter $c(M_\mathrm{vir},z)$ according to the model introduced above.

\begin{figure}
\centering
\includegraphics[scale=0.45]{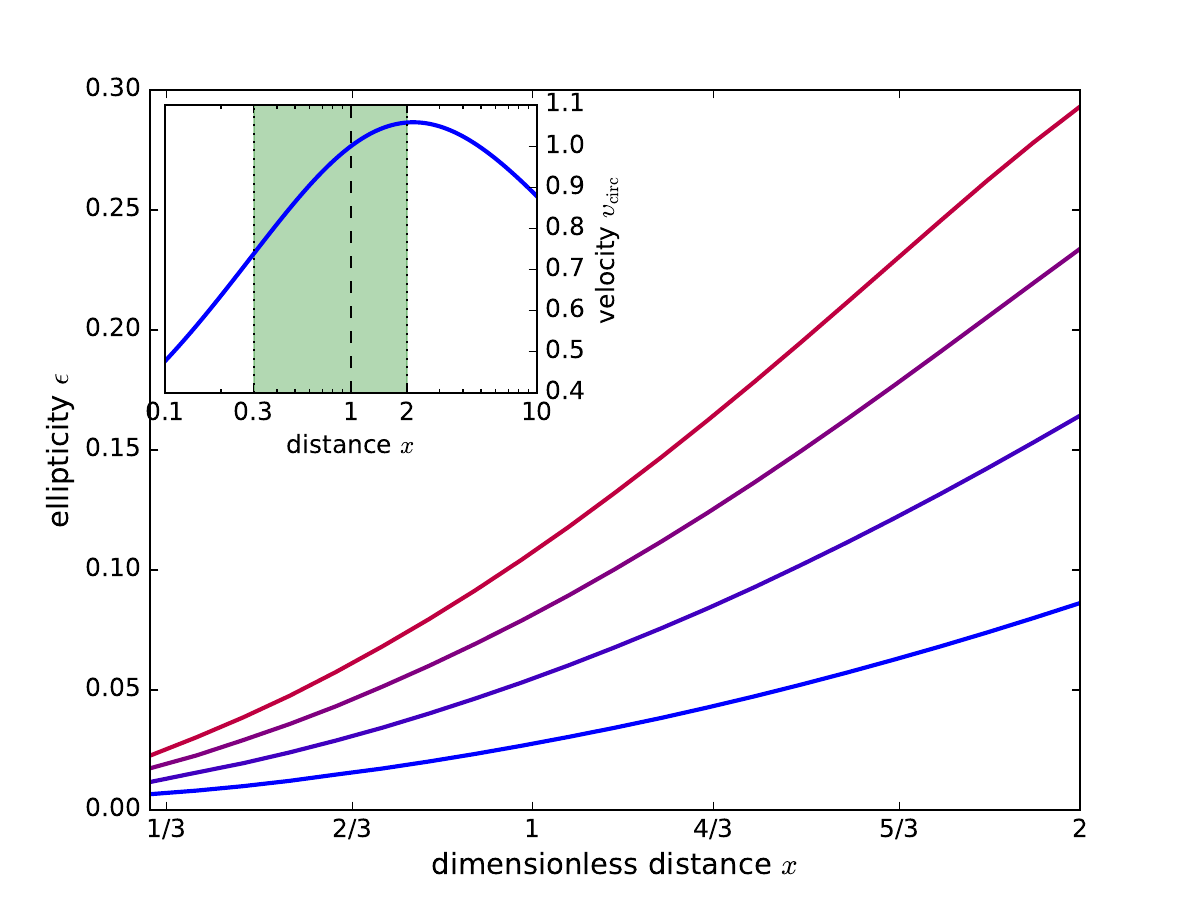}
\caption{Scaling of the ellipticity $\epsilon$ with orbital radius given in dimensionless units where $r = r_s$ corresponds to $x = 1$, for a fixed tidal shear field $\Omega^2 = 0.01, 0.02, 0.03, 0.04$ (from bottom to top). The inset shows the dimensionless rotation curve $\upsilon_\mathrm{circ}(x)$ for the range of radii $x=0.3\ldots2$ considered here.}
\label{fig_ellipticity_scaling}
\end{figure}

It is quite important to realise that there seems to be a fundamental difference in the linear alignment of elliptical and spiral galaxies: While the reaction of an elliptical galaxy as a virialised sphere to an external field is homologous, the induced ellipticity of the orbit of a star is a function of radius, as illustrated by Fig.~\ref{fig_ellipticity_scaling}. There is in fact a strong dependence of ellipticity on radius, depicted in the range $x = r/r_s = 1/3\ldots2$ caused by the fact that stars on more distant orbits are less tightly bound and therefore more susceptible to tidal distortion. We chose specifically this interval of radii to cover the range between the S{\'e}rsic-radius on small scales up to the edge of the galactic disc, where the circular velocity reaches the highest values, while emphasising that at even larger radii the tidal effect would be stronger.

From this point of view it becomes necessary to define a mean induced ellipticity, weighted with the exponential profile $\propto \exp(-r/R)$ with the S{\'e}rsic-radius $R$:
\begin{equation}
\bra\epsilon\ket = \frac{\int 2\pi r\dd r\:\epsilon(r)\exp(-r/R)}{\int 2\pi r\dd r\:\exp(-r/R)},
\end{equation}
which leads with an assumed exponential dependence $\epsilon(r) \propto \exp(+r/Q)$ suggested by Fig.~\ref{fig_ellipticity_scaling} with a scale $Q$ to
\begin{equation}
\bra\epsilon\ket = \frac{1}{\left(1-R/Q\right)^2}
\label{eqn_avg_ellipticty}
\end{equation}
which can in principle lead to large numbers if $Q\simeq R$, as the integral diverges for this choice. Given the numerical results of Fig.~\ref{fig_ellipticity_scaling}, the averaging will yield larger values for $\bra\epsilon\ket$ compared to the value for $\epsilon$ at $x=1$ by a few: $Q$ assumes values around unity, so that $R/Q\simeq 1/3$, and consequently, $\bra\epsilon\ket\simeq 9/4 = 2.25$. Looking back at eqn.~(\ref{eqn_fundamental_scaling}), this number corrects the expectation for $\epsilon_\mathrm{sp}$ in relation to $\epsilon_\mathrm{el}$ from $1/c(M_\mathrm{vir},z)^2\simeq 0.04$ to about 0.1, in agreement with the results by \citet{Hilbert:2016ylf} found in a fluid-mechanical simulation of galaxy formation and evolution in terms of the relative amplitudes of the intrinsic alignment spectra separated by early- and late-type galaxies. More recently, simulations of \citet{Jagvaral:2022zto} find alignments of disc galaxies that are not as weak as those reported by \citet{Hilbert:2016ylf}, which our model can accommodate by extending the radial averaging to larger distances, with a stronger response of the orbits to perturbing tidal fields. In summary, our analytical model does in fact explain ellipticities of disc galaxies that scale linearly with tidal field, and predicts the correct magnitude of the alignment. In the following, we work with a relative alignment parameter of $4$, intermediate between \citet{Hilbert:2016ylf} and \citet{Jagvaral:2022zto}.

\subsubsection{Tidal effect on moments of the brightness distribution}
Ellipticity as the actual observable pertains to the size of luminous component of the galaxy and the brightness profile $\rho_0(r)$, for which we assumed in previous works a S{\'e}rsic profile with characteristic radius $R$. The perturbation $\Delta Q_{ab}$ of the brightness moments $Q_{ab}$ of an elliptical galaxy can be computed with
\begin{equation}
\Delta Q_{ab} =
\frac{1}{2\sigma_\mathrm{vir}^2}\frac{\int\dd^2 r\:\rho_0(r)\: r_a r_b\: r_i r_j}{\int\dd^2r\:\rho_{0}(r)}\:
\partial^2_{ij}\Phi =
S_{abij}\:\partial^2_{ij}\Phi.
\end{equation}
From the second moments $Q_{ij}$ of the brightness distribution one defines the complex ellipticity $\epsilon$,
\begin{equation}
\epsilon \simeq
\frac{\Delta Q_{xx}-\Delta Q_{yy}}{Q_{xx}+Q_{yy}} + 2\ci\frac{\Delta Q_{xy}}{Q_{xx}+Q_{yy}}
\end{equation}
which directly reflects magnitude and orientation of the tidal shear field $\partial^2_{ij}\Phi$, with the susceptibility $S_{abij}$ as the constant of proportionality: It includes the virial velocity $\sigma_\mathrm{vir}^2$, and the details of the brightness distribution $\rho_0$.   For an elliptical, $S_{abij}$ is fundamentally an area of size $R^2$ for a S{\'e}rsic profile with scale $R$, and a numerical factor resulting from the integration that depends strongly on S{\'e}rsic-index $n$ \citep{Ghosh:2020zfa, Giesel:2021ikc, Giesel:2022org}: Typical numerical prefactors to $R^2$ resulting from the integration of the S{\'e}rsic-profile are $S\simeq 10$ for exponential profiles with $n=1$ up to $S\simeq10^3$ for de Vaucouleurs-profiles with $n=4$. The reaction of a galactic disc to a tidal shear field can in principle be encapsulated by the same relation, with $\upsilon_\mathrm{circ}^2$ replacing $\sigma_\mathrm{vir}^2$, and absorbing the averaging in eqn.~(\ref{eqn_avg_ellipticty}) into $S_{abij}$.

\subsubsection{Consistency relations}
For spiral galaxies we expect an identical argument to hold, with the choice of an exponential profile typical for galactic discs, and a radius-averaged ellipticity, as deduced in the previous chapter. But there are two major differences: Tidal alignments of orbits in the galactic disc are strongly radius-dependent, necessitating  the averaging process outlined in eqn.~(\ref{eqn_avg_ellipticty}). Secondly, the squared S{\'e}rsic-radius, which is a sensible choice for the alignment parameter of elliptical galaxies as it reproduces the alignment amplitude and corresponds to the physical picture of a tidally distorted virialised object, is remarkably  similar for spiral and elliptical galaxies of similar mass. Therefore, it cannot be the alignment parameter for a disc. If one stipulates that $SR^2 = r_s^2$ for disc galaxies and $SR^2 = r_\mathrm{vir}^2$ for elliptical galaxies, the missing link between the dynamical properties of the dark matter halo (encapsulated in $r_s$ and $r_\mathrm{vir}$) and the light distribution of the luminous component (described by $R$ and $n$) can potentially be established: Numerically, this would imply for a Milky Way-sized halo of $\sim 10^{12}M_\odot/h$ with a virial radius of $\sim 100~\mathrm{kpc}/h$ \citep{Dehnen:2006cm}, an NFW-scale radius $r_s\sim 10~\mathrm{kpc}/h$ \citep{Lin:2019yux} and a typical S{\'e}rsic-radius of a few kpc \citep{Goodwin:1997ys,1998Obs...118..201G}, which would not be too dissimilar for an elliptical of the same mass.

\subsubsection{Observational consequences}
Through our scaling argument, one would assume the reaction of disc galaxies to be very similar compared to that of elliptical galaxies. In addition, because of the model's linearity, there would be $(i)$ a $GI$-type cross-correlation between the intrinsic shapes of disc galaxies and gravitational lensing with the characteristic negative amplitude, as well as $(ii)$ an $II$-type correlation between the intrinsic shapes of disc and elliptical galaxies, with alignment parameters that follow consistently from the linear tidal interaction model. Additionally, $(iii)$ one can speculate about an effect of $\Delta\Phi$ as the trace of the tidal field on the size of the galactic disc, giving rise to a size-density correlation analogous to that in ellipticals: Galactic discs in low-density environments should be larger; an effect which we elaborate on in the Appendix. Finally, $(iv)$ unlike angular momentum-based alignment models, the linear alignment model would not give rise to $B$-modes in the ellipticity field and $(v)$ there should be a nonzero bispectrum of the ellipticity field at tree-level first order perturbation theory, analogous to gravitational lensing. All these points show that linear alignments for spiral galaxies are markedly different compared to quadratic alignments, as they would result from an angular momentum-based model.

\subsubsection{Tidal fields in the large-scale structure}
With these considerations, the magnitude of intrinsic alignments tidally generated by the large-scale structure can be estimated: The magnitude of the perturbing tidal field is $\Omega^2$, which needs to be small compared to the strength of the potential binding the stars to the galaxy: In fact, the centrifugal balance condition implies
\begin{equation}
\omega^2 = \frac{GM_\mathrm{vir}}{r_s^3}\frac{B}{A}
\end{equation}
which is effectively the third Kepler-law modified by a numerical factor $B/A\neq 1$. Then, with $M_\mathrm{vir} = 4\pi/3\:r_\mathrm{vir}^3\Delta\Omega_m\rho_\mathrm{crit}$ with the virial overdensity $\Delta \simeq200$ and the relation $r_s = r_\mathrm{vir}/c(M_\mathrm{vir},z)$ for NFW-haloes one arrives at:
\begin{equation}
\omega^2 = \frac{c(M_\mathrm{vir},z)^3}{2}\frac{B}{A}\Delta\Omega_m H_0^2
\end{equation}
such that the orbital time scale $1/\omega$ is shorter by a factor of $c^3\Delta\Omega_m$ compared to the Hubble time scale $1/H_0$, where $\Delta$ and $\Omega_m$ are universal, but $c(M_\mathrm{vir},z)$ remains a weak function of $M_\mathrm{vir}$ and $z$. A typical perturbation of the density field in the large-scale structure would have an amplitude of $\delta\simeq1$, resulting in a perturbation of the potential following the Poisson-equation $\Delta\Phi = 3/2\:\Omega_mH^2$, which gives immediately a value of $\Omega^2 = 3/2\:\Omega_m H_0^2$ one can expect for the strength of the tidal field. Then, the perturbations are necessarily weak,
\begin{equation}
\frac{\omega^2}{\Omega^2} = \frac{c(M_\mathrm{vir},z)^3}{3}\frac{B}{A}\Delta \gg 1,
\end{equation}
driven by the fact that $\Delta$ and $c(M_\mathrm{vir},z)^3$ are much larger than one. Reiterating this reasoning for elliptical galaxies would start at virial equilibrium instead of centrifugal equilibirum, $\omega^2 = \sigma_\mathrm{vir}^2/r_\mathrm{vir}^2 = GM_\mathrm{vir}/r_\mathrm{vir}^3$ such that
\begin{equation}
\omega^2 = \frac{1}{2}\Delta\Omega_mH_0^2
\end{equation}
and the tidal field exerted by the large-scale structure is again weak, $\omega^2/\Omega^2 = \Delta/3\gg 1$. On both cases, the galaxies are sufficiently strongly bound such that typical tidal fields are weak perturbations to their own gravitational fields. We suspect that the inverse density scale introduced in the common parameterisation of the linear alignment model for elliptical galaxies \citep{Hirata:2004gc, Brown:2000gt,Blazek:2017wbz,Blazek:2011xq,Blazek:2015lfa}
\begin{equation}
\epsilon \simeq A_I\delta
\quad\mathrm{with}\quad
A_I = -a_1 \tilde{C}_1 \frac{\rho_\mathrm{crit}\Omega_m}{D_+(z)}
\end{equation}
with $\tilde{C}_1$ fixed to $5\times10^{-14}(\mathrm{Mpc}/h)^3/(M_\odot/h)$ effectively reproduces with the background density $\rho_\mathrm{crit}\Omega_m$ the virial overdensity $\Delta$, as it determines the ratio between internal and external gravitational fields. This has indeed been proven to be true and mathematically derived previously in \citep{Hirata:2004gc}.

\subsection{Tidal fields in 3 dimensions}
\begin{figure}
\centering
\includegraphics[scale=0.45]{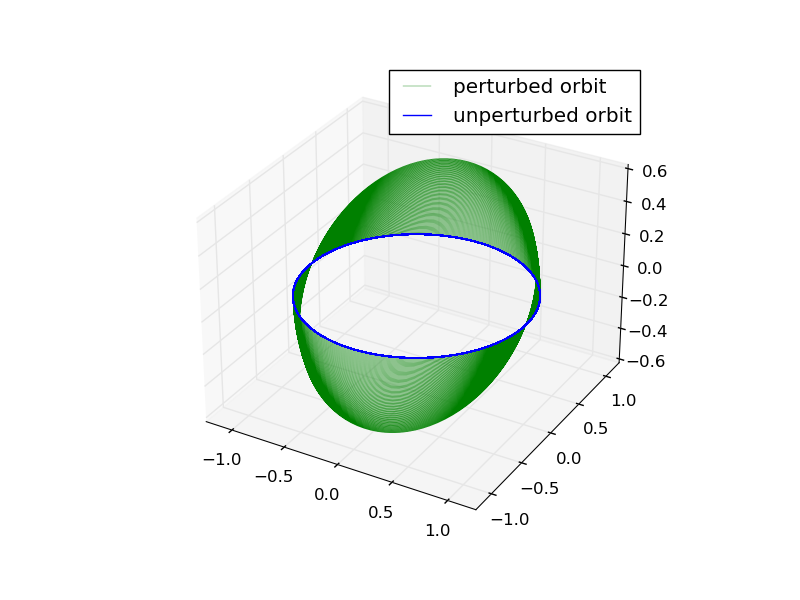}
\includegraphics[scale=0.45]{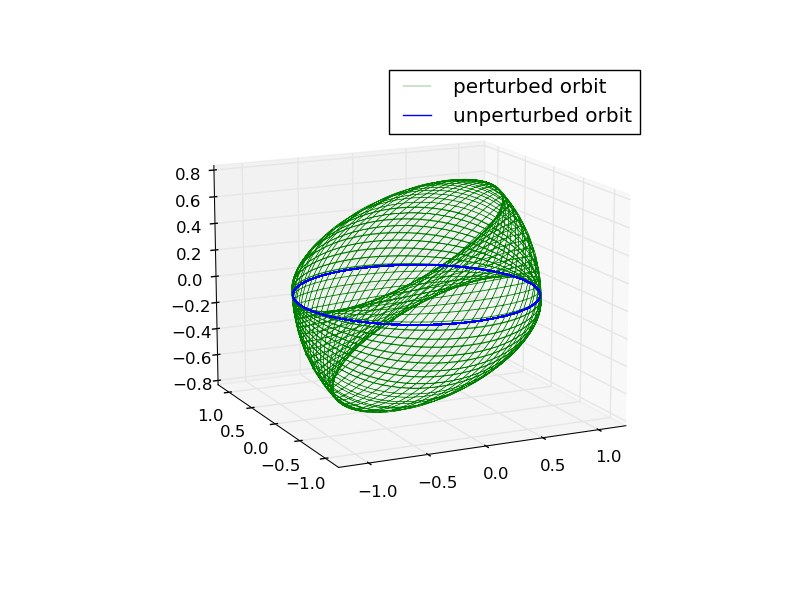}
\caption{Perturbed orbits (green) in comparison to unperturbed circular, planar orbits (blue) in 3 dimensions, for $\lambda^{(4)}$ (top panel; $\lambda^{(6)}$ would show the same alignment rotated by 90 degrees) and $\lambda^{(8)} + \lambda^{(4)}$ (bottom panel, the linear combination was chosen as the effect of $\lambda^{(8)}$ alone would be in the plane of the disc).}
\label{fig_3d}
\end{figure}

Up to this point, the perturbing tidal shear field was restricted to the galactic plane: Generalising this situation leads to a perturbing potential
\begin{equation}
\Phi = \frac{\Omega^2}{2}\lambda^{(n)}_{ij}x_ix_j
\quad\rightarrow\quad
\partial_i\partial_j\Phi = \Omega^2\lambda^{(n)}_{ij}
\end{equation}
where a suitable basis $\lambda^{(n)}_{ij}$ could be provided by the symmetric subset of Gell-Mann matrices (as a generalisation of the Pauli-matrices used in the analogous expansion in eqns.~\ref{eqn_expansion_rirj} and~\ref{eqn_expansion_phiij}), which form together with the unit matrix a basis of all tidal shear fields $\partial_i\partial_j\Phi$. There are six symmetric Gell-Mann matrices, and the effect of two of them, specifically $\lambda^{(1)}$ and $\lambda^{(3)}$, was already investigated above, as they contain the Pauli-matrices in their top left $2\times2$ submatrix:
\begin{equation}
\lambda^{(1)} =
\left(
\begin{array}{ccc}
0 & 1 & 0\\ 1 & 0 & 0\\ 0 & 0 & 0
\end{array}
\right),
\quad
\lambda^{(3)} =
\left(
\begin{array}{ccc}
1 & 0 & 0\\ 0 & -1 & 0\\ 0 & 0 & 0
\end{array}
\right).
\end{equation}
The unit matrix, which is likewise an element of the Gell-Mann matrices, would cause an isotropic change in size of the galactic disc (c.f. Appendix). The two Gell-Mann matrices $\lambda^{(4)}$ and $\lambda^{(6)}$,
\begin{equation}
\lambda^{(4)} =
\left(
\begin{array}{ccc}
0 & 0 & 1\\ 0 & 0 & 0\\ 1 & 0 & 0
\end{array}
\right),
\quad
\lambda^{(6)} =
\left(
\begin{array}{ccc}
0 & 0 & 0\\ 0 & 0 & 1\\ 0 & 1 & 0
\end{array}
\right),
\end{equation}
are able to tilt the orbital plane of the disc, as exemplified by Fig.~\ref{fig_3d}. This effect can either increase or decrease the projected ellipticity, depending on the unperturbed orientation of the galactic disc. It evolves dynamically, as the effect of the tidal gravitational field perpendicular to the galactic disc is cumulative. Finally, the Gell-Mann matrix $\lambda^{(8)}$,
\begin{equation}
\lambda^{(8)} =
\frac{1}{\sqrt{3}}
\left(
\begin{array}{ccc}
1 & 0 & 0\\ 0 & 1 & 0\\ 0 & 0 & -2
\end{array}
\right),
\end{equation}
combines the effect of changing the size of the disc in an isotropic fashion while at the same time causing oscillations above and below the unperturbed orbital plane. Because the orbits are not closed in the perturbed potential anymore, the orbital plane will undergo a precession motion which effectively leads to an isotropisation of the system and a more spherical appearance with low ellipticity as shown in Fig.~\ref{fig_3d}.

\section{Ellipticity spectra}\label{sect_ellipticity}
The primary observable of weak lensing surveys are (tomographic) ellipticity spectra, which map out line of-sight-integrated fluctuations of the tidal gravitational field in a wide redshift range reaching out beyond unity. As a subdominant contribution to the lensing-induced ellipticity spectra there are intrinsic alignments of elliptical galaxies, which reflect redshift-distribution averaged correlations of tidal shear fields in the respective redshift bin. With the results from the previous chapters, namely that disc galaxies follow the same linear alignment model as ellipticals albeit with a smaller alignment parameter, one can relate their intrinsic ellipticity spectra by an overall scaling. Numerically, we follow the analysis outlined by \citet{Ghosh:2020zfa}, to which we refer the reader.

\begin{figure}
\centering
\includegraphics[scale=0.4]{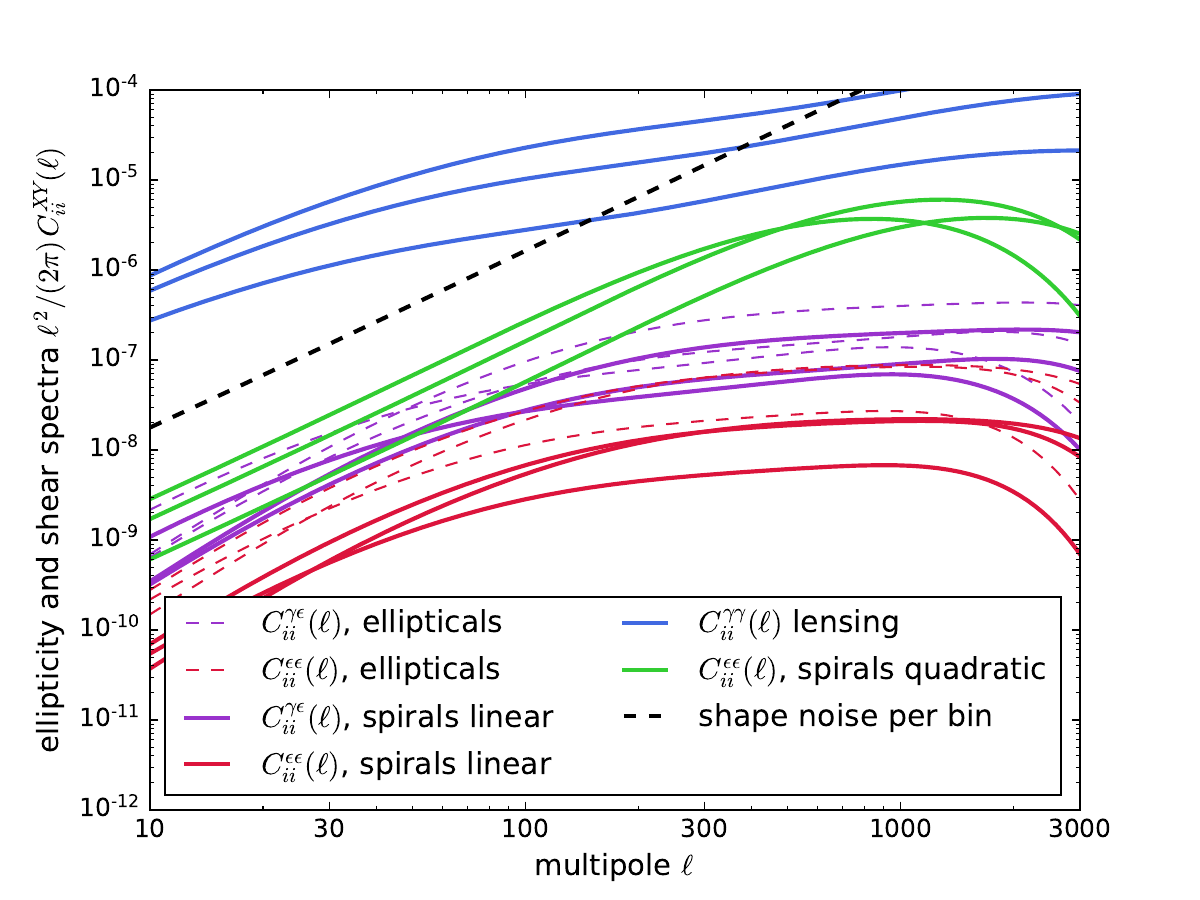}
\caption{Angular ellipticity spectra for a simplified three-bin tomographic survey with Euclid's redshift distribution, comparing weak gravitational lensing with intrinsic alignments of elliptical galaxies, and in particular with alignments of spiral galaxies according to a quadratic and to a linear alignment model. We assume that spirals make up two thirds of the galaxy population, and ellipticals the remaining third.}
\label{fig_spectrum}
\end{figure}

Angular ellipticity spectra for lensing $C_{ij}^{\gamma\gamma}(\ell)$, for intrinsic alignments $C_{ij}^{\epsilon\epsilon}(\ell)$ and for their cross-correlation $C_{ij}^{\gamma\epsilon}(\ell)$ are shown in Fig.~\ref{fig_spectrum}, for an Euclid-like survey with median redshift of 0.9. For clearer visualisation, we choose to show a simplified setup with only three bins, with equal fractions of the total galaxy population. While there are no cross-correlations across different redshifts bins in intrinsic alignments, cross-correlations of lensing do exist, but are excluded in the plots in order not to overcrowd the figure. Parallel to the $GI$ and $II$-spectra for elliptical galaxies we show the corresponding spectra of spiral galaxies as well, both weighted by the relative fraction of elliptical galaxies (making up roughly a third of the galaxy population) and spiral galaxies (accounting for the remaining two thirds). The shape noise term assumes a standard value of $\sigma_\epsilon=0.4$ for an individual galaxy, scaled down with $n = 4.727\times10^8$ galaxies per steradian.

\begin{figure}
\centering
\includegraphics[scale=0.4]{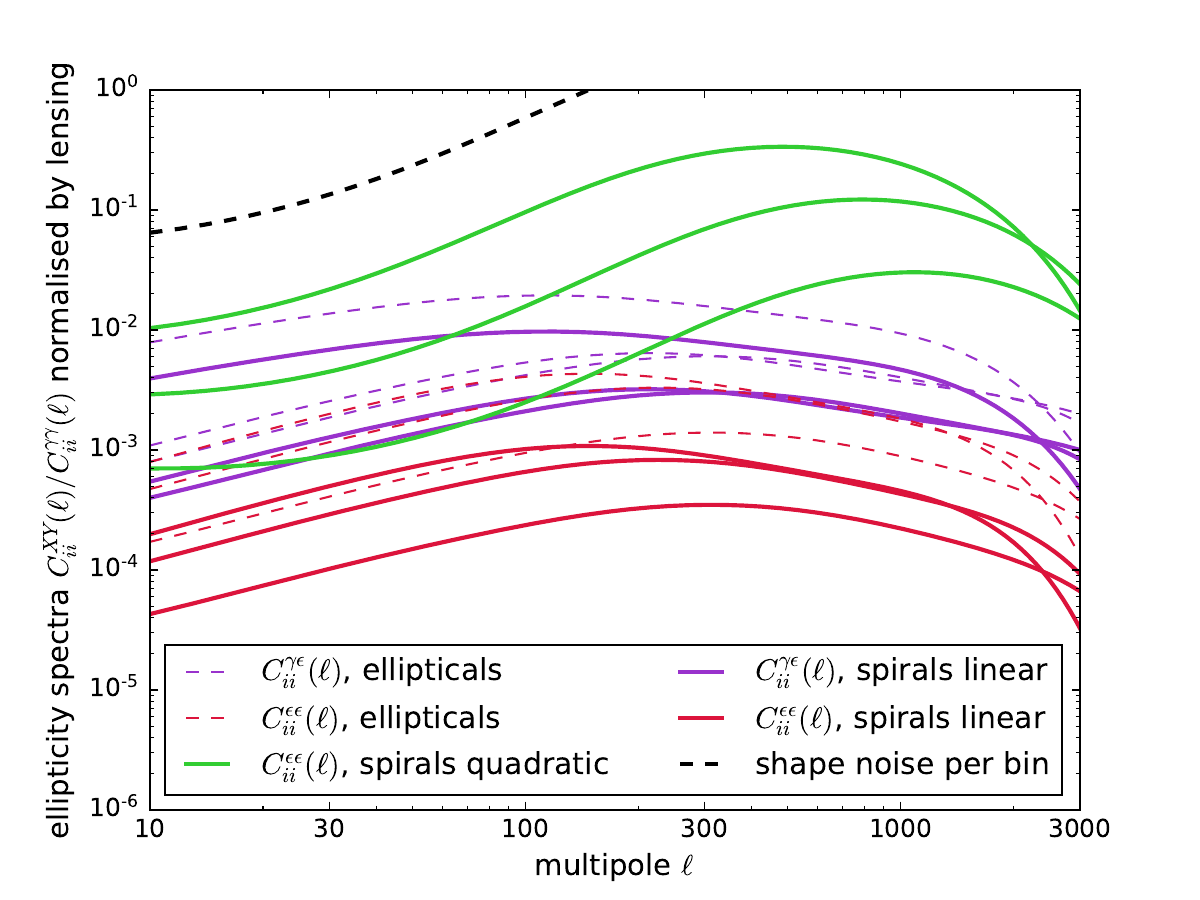}
\caption{Ratios between the angular ellipticity spectra of intrinsic alignments and weak gravitational lensing analogous to Fig.~\ref{fig_spectrum}, for three bins and for $i=j$. It is clear that there is a large overprediction in case of the quadratic alignment model for spiral galaxies.}
\label{fig_ratio}
\end{figure}

The ratios $C^{\epsilon\epsilon}_{ij}(\ell)/C^{\gamma\gamma}_{ij}(\ell)$ and $C^{\gamma\epsilon}_{ij}(\ell)/C^{\gamma\gamma}_{ij}(\ell)$ between the intrinsic alignment spectra and the weak lensing spectrum are shown in Fig.~\ref{fig_ratio}, again for only three bins and for the choice $i=j$. Most importantly, one can see by how much an angular-momentum based, quadratic alignment model would overpredict the effect \citep{Schafer:2015qfx}, and how the linear alignment model for spiral galaxies evades detection in earlier data sets \citep{Heymans:2013fya}. But most importantly, the dynamical model suggests that the alignment amplitudes of elliptical and spiral galaxies should be related, as the alignment parameter reflects effectively the area of the luminous component of the systems in both cases. This suggests the question whether marked correlation functions which combine intrinsic ellipticity correlations with galaxy properties such as S{\'e}rsic-indices or S{\'e}rsic-radii can isolate more information about the alignment mechanisms.

We start with the numbers derived by \citet{Ghosh:2020zfa}, who derive estimates on the observability of ellipticity-lensing correlation in aligned ellipticals in an Euclid-like survey. For ellipticals making up $q=1/3$ of the galaxy population and an alignment parameter $D_{IA} = -(10^{-5}~\mathrm{Mpc}/h)^2$, an Euclid-like redshift distribution with a median redshift of $0.9$, 5 tomographic bins and integrating up all multipoles from $\ell = 10$ to $\ell = 3000$ yields a statistical significance $\Sigma/\sqrt{f_\mathrm{sky}}\simeq 10\ldots15$, where $\Sigma$ is the signal-to-noise ratio and $f_\mathrm{sky}$ is the sky coverage of the survey, including effects on nonlinear structure formation (and correspondingly higher tidal shear fields) on small scales. Assuming that the measurement is limited by shape noise and by cosmic variance from weak gravitational lensing, one can use the proportionality of the spectra to scale down the significance for ellipticity correlations of disc galaxies: Relative to the $GI$-spectrum for ellipticals which is measurable at $\Sigma/\sqrt{f_\mathrm{sky}}\simeq 15$, the $GI$-spectrum for spirals should generate a signal at a significance lower by $\sqrt{(1-q)/q/\alpha} \simeq 0.7$. We assume the ratio of alignment parameters $\alpha$ a value of $4$, intermediate between \citet{Hilbert:2016ylf} and \citet{Jagvaral:2022zto}. Relative to the $II$-spectrum of elliptical galaxies can be measured at a signficiance of about $\Sigma/\sqrt{f_\mathrm{sky}}\simeq 10$, the $II$-spectrum of spirals alone should generate a significance lower by $(1-q)/q/\alpha \simeq 0.5$, and the $II$-type cross-correlation between spirals and ellipticals would be at significances lower by $\sqrt{(1-q)/\alpha} \simeq 0.4$. This effectively puts all spectra within reach of an Euclid-type survey.

\section{Summary and discussion}\label{sect_summary}
The subject of this paper is the construction of a linear alignment model for disc galaxies that conforms with the results of simulations of galaxy formation and with observations of intrinsic alignments in weak lensing data. In order to agree with observations, alignments of disc galaxies must be weaker than in elliptical galaxies, and simulations do in fact suggest a weak alignment at linear instead of quadratic order, with the interesting consequence that $GI$-type alignments should exist in disc galaxies, very much in contrast to the predictions by quadratic, angular-momentum based alignment models.

\begin{enumerate}
\item{The first step was a perturbative calculation, with an associated energy consideration, to derive the perturbative effect of a quadrupolar tidal gravitational field on a circular orbit. We showed that the ellipticity reflects the ratio between the squared orbital frequency of the unperturbed orbit and the magnitude of the tidal gravitational field. This model produces naturally an ellipticity which is proportional to the magnitude of the tidal field. Being more specific, we then set up a direct simulation of the orbit of a star inside the gravitational potential of an NFW-halo, tidally perturbed again by a quadrupolar field. This direct simulation confirmed the perturbative results. In addition, we could show that the induced ellipticity depends on orbital radius, as stars at higher radii are less bound and are more susceptible to tidal perturbation.}
\item{With the alignment of both elliptical and spiral galaxies following a linear alignment model and both being determined by the external tidal gravitational field, the question of how the relative alignments are determined by galaxy properties arises naturally. Elliptical galaxies react in a homologous way to tidal fields, whereas the tidally-induced change in shape of spirals depends on radius. The linear alignment model would now stipulate that the alignment of spirals is weaker as the discs are smaller than the virial radius of the host halo. At the NFW-scale radius, the ellipticity of a galactic disc is $c(M_\mathrm{vir},z)^2$ times smaller than that of an elliptical, with $c(M_\mathrm{vir},z)$ being the NFW-concentration parameter. Averaging the ellipticity over all radii, weighted with an exponential brightness distribution, alleviates this number and sets the alignments strength of spirals to be about an order of magnitude smaller than that of ellipticals.}
\item{The physical picture behind linear alignments of disc galaxies suggests a number of observational relations that could be tested with data: The alignment should be stronger with higher S{\'e}rsic-index, and higher for systems with larger discs. There should be a scaling of the alignment with the mass of the system, driven by the change in NFW-concentration parameter $c(M_\mathrm{vir},z)$, too: As low-mass systems tend to be more centrally concentrated, their alignment should be weaker. It can also be checked whether empirically introduced laws to allow for redshift evolution can be replaced by the known evolution of the NFW-concentration parameter $c(M_\mathrm{vir},z)$: Here, the alignment would be highest for high $z$ and decrease towards the current epoch. To what extent evolution of alignments with redshift and their scaling with mass are driven by the dependence of the NFW-concentration parameter and what additional astrophysical effects could play a role, would be an interesting question for surveys and for simulations.}
\item{A further interesting consequence of the linear alignment model are changes in shape: Both ellipticals and spirals should tend to be larger in low-density regions. Flexion-like distortions from higher-order tidal shear fields on galactic discs should exist but would be very small, given the result that the effect induced by $\partial^3\Phi$ on ellipticals as investigated in \citet{Giesel:2021ikc} is already small. But physically, these distortions would be able to warp a galactic disc, which would constitute a link between alignments and disc dynamics.}
\item{Ellipticity spectra resulting from the linear alignment model naturally resemble those of elliptical galaxies as they are related just by a constant prefactor. With their smaller amplitudes, ellipticity correlations of spiral galaxies evade constraints on alignments derived from current weak lensing data and do not produce signals that can be measured at reasonable significances. More important, though, are the much smaller amplitudes of the linear alignment model in spiral galaxies compared to angular-momentum based models: Typically, the spectra are smaller by more than an order of magnitude at high multipoles. Additionally, the phenomenology of the spectra is quite different: In contrast to the quadratic, angular-momentum based alignment model which predicts no cross-correlation between the shapes of elliptical and spiral galaxies, the linear alignment model would predict cross-correlations between the shapes of the two galaxy types in a very natural way. Moreover, the linear alignment model would produce pure $E$-mode correlations in the ellipticity field, which would apply likewise for spiral and elliptical galaxies. Again, the quadratic model would be markedly different, with a high amplitude $B$-mode correlation being predicted. Lastly, one would expect a nonzero cross-correlation between intrinsic shapes of both ellipticals and spirals with gravitational lensing shear, with the typical anti-correlations present in both galaxy types. Linking tidal alignment to galaxy properties one should expect a dependence on S{\'e}rsic-index as more centrally concentrated objects would show weaker alignments. Estimates suggest that all intrinsic ellipticity spectra are within reach of Euclid. As outlined in \cite{Euclid:2022vtv}, S\'ersic profiles will play a crucial role in distinguishing galaxy morphologies in Euclid.}
\end{enumerate}

Spiral galaxies regularly have galactic bulges at their centres, which show analogous alignments in simulations by \citet{Jagvaral:2022zto}, up to the fact that they tend to have rather high S{\'e}rsic indices close to $n = 4$, are rather strongly bound with high velocity dispersions $\sigma$ and are rather compact with a small S{\'e}rsic radii $R$; numbers of $\sigma = 140~\mathrm{km}/\mathrm{s}$ and $R = 1/2~\mathrm{kpc}$ have been reported for Milky Way \citep{Valenti_2018}. Their amplitude seems to be similar to that of elliptical galaxies, and is much stronger than that of the host disc galaxy. Whether this is a consequence of the homologous distortion of virial systems remains to be investigated. We intend to develop the topic of tidally distorted galactic discs and the resulting ellipticity correlations further by considering the full 3-dimensional tidal field, which would effectively bridge between the concept of disc warps and intrinsic alignments, or perhaps even link alignments with tidally-induced bar formation in spiral galaxies.

\section*{Acknowledgements}
We would like to thank Robert Reischke and Joerg Jaeckel for their thoughtful remarks.

\paragraph{Funding information}
BG would also like to acknowledge financial support from the DST-INSPIRE Faculty fellowship (Grant no. DST/INSPIRE/04/2020/001534) and the DAAD Research Stays for University Academics and Scientists Fellowship (Ref No. 91879665), and Heidelberg University for hosting during the research visit. ESG acknowledges support from the German National Academic Foundation.

\section*{Appendix: Isotropic increase in size}\label{sect_size}
\begin{figure}
\centering
\includegraphics[scale=0.45]{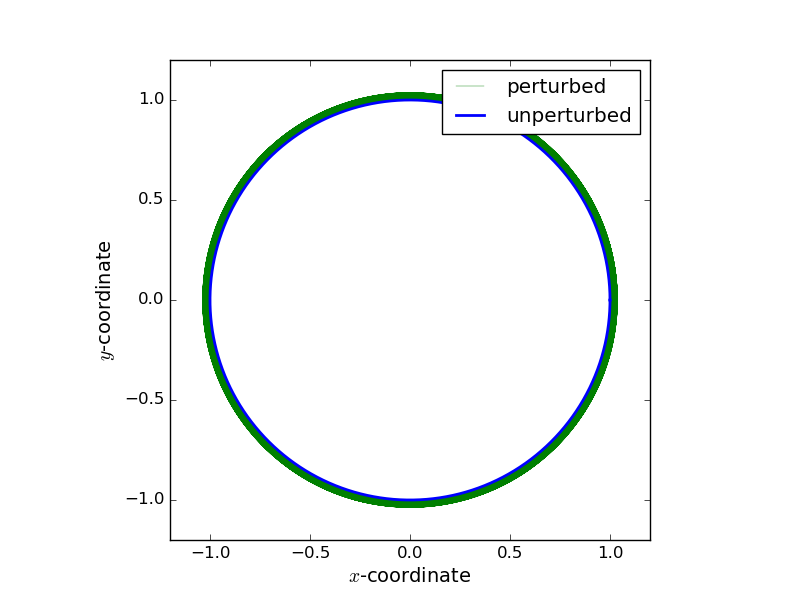}
\caption{Perturbed orbits (green) in comparison to unperturbed circular orbits (blue) at $r=r_s$ in the potential of an NFW-halo with an isotropic tidal field $\Omega^2(x^2+y^2)$ giving rise to size change of the disc analogous to the lensing convergence $\kappa$. The magnitude of the perturbing field is again $\Omega^2 = 0.04$ in dimensionless units.}
\label{fig_isotropic_2d}
\end{figure}

Fig.~\ref{fig_isotropic_2d} shows the effect of an isotropic tidal field $\Delta\Phi$ to the orbit of a star at $r = r_s$. As expected, there is a change in size of the galactic disc, analogous to the effect on elliptical galaxies \citep{Ghosh:2020zfa} or, equivalently, to the effect of lensing convergence \citep{bartelmann_weak_2001}. In both cases, galaxies would appear larger in low-density environments and smaller in high-density environments. Visually, the effect of an isotropic tidal field on the size of the disc seems to be smaller compared to the shear depicted in Fig.~\ref{fig_perturbed}, but one perceives only half of the effect as both semi-axes are changed by the same factor $1+\Omega^2/\omega^2/2$ and unlike in the case of ellipticity, $1+\Omega^2/\omega^2/2$ relative to $1-\Omega^2/\omega^2/2$. The same argument as for disc-averaged ellipticities would apply here as well, as less tightly bound stars at larger radii would experience a stronger tidal enlargement of their orbital radii. In consequence, one should expect a size alignment effect with an equal amplitude as the ellipticity alignment effect, in analogy to the result for size alignments in elliptical galaxies \citep{Ghosh:2020zfa}. Correlations between size and alignments have also been highlighted previously in literature \citep{Vlah:2019byq,Ciarlariello:2014qva,Ciarlariello:2016hxr,Johnston:2022nbv}. As the size-alignments for elliptical galaxies are difficult to measure even for surveys such as Euclid's, it is doubtful if the analogous measurement for disc galaxies could yield a statistically significant detection, extrapolating the results by \citet{Ghosh:2020zfa}. Repeating the estimates in Sect.~\ref{sect_ellipticity} in conjunction with the results of \citet{Ghosh:2020zfa} yields the result that correlations between size perturbations and weak lensing convergence are likewise difficult to measure with data from an Euclid-like survey.

\bibliography{references.bib}

\begin{thebibliography}{57}
\expandafter\ifx\csname natexlab\endcsname\relax\def\natexlab#1{#1}\fi
\expandafter\ifx\csname bibnamefont\endcsname\relax
  \def\bibnamefont#1{#1}\fi
\expandafter\ifx\csname bibfnamefont\endcsname\relax
  \def\bibfnamefont#1{#1}\fi
\expandafter\ifx\csname citenamefont\endcsname\relax
  \def\citenamefont#1{#1}\fi
\expandafter\ifx\csname url\endcsname\relax
  \def\url#1{\texttt{#1}}\fi
\expandafter\ifx\csname urlprefix\endcsname\relax\def\urlprefix{URL }\fi
\providecommand{\bibinfo}[2]{#2}
\providecommand{\eprint}[2][]{\url{#2}}

\bibitem[{\citenamefont{Hirata and Seljak}(2004)}]{Hirata:2004gc}
\bibinfo{author}{\bibfnamefont{C.~M.} \bibnamefont{Hirata}} \bibnamefont{and} \bibinfo{author}{\bibfnamefont{U.}~\bibnamefont{Seljak}}, \bibinfo{journal}{Phys. Rev. D} \textbf{\bibinfo{volume}{70}}, \bibinfo{pages}{063526} (\bibinfo{year}{2004}), \bibinfo{note}{[Erratum: Phys.Rev.D 82, 049901 (2010)]}, \eprint{astro-ph/0406275}.

\bibitem[{\citenamefont{Hirata et~al.}(2007)\citenamefont{Hirata, Mandelbaum, Ishak, Seljak, Nichol, Pimbblet, Ross, and Wake}}]{Hirata:2007np}
\bibinfo{author}{\bibfnamefont{C.~M.} \bibnamefont{Hirata}}, \bibinfo{author}{\bibfnamefont{R.}~\bibnamefont{Mandelbaum}}, \bibinfo{author}{\bibfnamefont{M.}~\bibnamefont{Ishak}}, \bibinfo{author}{\bibfnamefont{U.}~\bibnamefont{Seljak}}, \bibinfo{author}{\bibfnamefont{R.}~\bibnamefont{Nichol}}, \bibinfo{author}{\bibfnamefont{K.~A.} \bibnamefont{Pimbblet}}, \bibinfo{author}{\bibfnamefont{N.~P.} \bibnamefont{Ross}}, \bibnamefont{and} \bibinfo{author}{\bibfnamefont{D.}~\bibnamefont{Wake}}, \bibinfo{journal}{Mon. Not. Roy. Astron. Soc.} \textbf{\bibinfo{volume}{381}}, \bibinfo{pages}{1197} (\bibinfo{year}{2007}), \eprint{astro-ph/0701671}.

\bibitem[{\citenamefont{Bridle and King}(2007)}]{Bridle:2007ft}
\bibinfo{author}{\bibfnamefont{S.}~\bibnamefont{Bridle}} \bibnamefont{and} \bibinfo{author}{\bibfnamefont{L.}~\bibnamefont{King}}, \bibinfo{journal}{New J. Phys.} \textbf{\bibinfo{volume}{9}}, \bibinfo{pages}{444} (\bibinfo{year}{2007}), \eprint{0705.0166}.

\bibitem[{\citenamefont{Crittenden et~al.}(2001)\citenamefont{Crittenden, Natarajan, Pen, and Theuns}}]{Crittenden:2000wi}
\bibinfo{author}{\bibfnamefont{R.~G.} \bibnamefont{Crittenden}}, \bibinfo{author}{\bibfnamefont{P.}~\bibnamefont{Natarajan}}, \bibinfo{author}{\bibfnamefont{U.-L.} \bibnamefont{Pen}}, \bibnamefont{and} \bibinfo{author}{\bibfnamefont{T.}~\bibnamefont{Theuns}}, \bibinfo{journal}{Astrophys. J.} \textbf{\bibinfo{volume}{559}}, \bibinfo{pages}{552} (\bibinfo{year}{2001}), \eprint{astro-ph/0009052}.

\bibitem[{\citenamefont{Schaefer and Merkel}(2012)}]{Schaefer:2011gt}
\bibinfo{author}{\bibfnamefont{B.~M.} \bibnamefont{Schaefer}} \bibnamefont{and} \bibinfo{author}{\bibfnamefont{P.}~\bibnamefont{Merkel}}, \bibinfo{journal}{Mon. Not. Roy. Astron. Soc.} \textbf{\bibinfo{volume}{421}}, \bibinfo{pages}{2751} (\bibinfo{year}{2012}), \eprint{1101.4584}.

\bibitem[{\citenamefont{Chisari et~al.}(2015)\citenamefont{Chisari, Codis, Laigle, Dubois, Pichon, Devriendt, Slyz, Miller, Gavazzi, and Benabed}}]{Chisari:2015qga}
\bibinfo{author}{\bibfnamefont{N.~E.} \bibnamefont{Chisari}}, \bibinfo{author}{\bibfnamefont{S.}~\bibnamefont{Codis}}, \bibinfo{author}{\bibfnamefont{C.}~\bibnamefont{Laigle}}, \bibinfo{author}{\bibfnamefont{Y.}~\bibnamefont{Dubois}}, \bibinfo{author}{\bibfnamefont{C.}~\bibnamefont{Pichon}}, \bibinfo{author}{\bibfnamefont{J.}~\bibnamefont{Devriendt}}, \bibinfo{author}{\bibfnamefont{A.}~\bibnamefont{Slyz}}, \bibinfo{author}{\bibfnamefont{L.}~\bibnamefont{Miller}}, \bibinfo{author}{\bibfnamefont{R.}~\bibnamefont{Gavazzi}}, \bibnamefont{and} \bibinfo{author}{\bibfnamefont{K.}~\bibnamefont{Benabed}}, \bibinfo{journal}{Mon. Not. Roy. Astron. Soc.} \textbf{\bibinfo{volume}{454}}, \bibinfo{pages}{2736} (\bibinfo{year}{2015}), \eprint{1507.07843}.

\bibitem[{\citenamefont{Rodriguez-Gomez et~al.}(2017)}]{Rodriguez-Gomez:2016jue}
\bibinfo{author}{\bibfnamefont{V.}~\bibnamefont{Rodriguez-Gomez}} \bibnamefont{et~al.}, \bibinfo{journal}{Mon. Not. Roy. Astron. Soc.} \textbf{\bibinfo{volume}{467}}, \bibinfo{pages}{3083} (\bibinfo{year}{2017}), \eprint{1609.09498}.

\bibitem[{\citenamefont{Schaefer}(2009)}]{Schaefer:2008xd}
\bibinfo{author}{\bibfnamefont{B.~M.} \bibnamefont{Schaefer}}, \bibinfo{journal}{Int. J. Mod. Phys. D} \textbf{\bibinfo{volume}{18}}, \bibinfo{pages}{173} (\bibinfo{year}{2009}), \eprint{0808.0203}.

\bibitem[{\citenamefont{Joachimi et~al.}(2015)}]{Joachimi:2015mma}
\bibinfo{author}{\bibfnamefont{B.}~\bibnamefont{Joachimi}} \bibnamefont{et~al.}, \bibinfo{journal}{Space Sci. Rev.} \textbf{\bibinfo{volume}{193}}, \bibinfo{pages}{1} (\bibinfo{year}{2015}), \eprint{1504.05456}.

\bibitem[{\citenamefont{Kiessling et~al.}(2015)}]{Kiessling:2015sma}
\bibinfo{author}{\bibfnamefont{A.}~\bibnamefont{Kiessling}} \bibnamefont{et~al.}, \bibinfo{journal}{Space Sci. Rev.} \textbf{\bibinfo{volume}{193}}, \bibinfo{pages}{67} (\bibinfo{year}{2015}), \bibinfo{note}{[Erratum: Space Sci.Rev. 193, 137 (2015)]}, \eprint{1504.05546}.

\bibitem[{\citenamefont{Kirk et~al.}(2015)}]{Kirk:2015nma}
\bibinfo{author}{\bibfnamefont{D.}~\bibnamefont{Kirk}} \bibnamefont{et~al.}, \bibinfo{journal}{Space Sci. Rev.} \textbf{\bibinfo{volume}{193}}, \bibinfo{pages}{139} (\bibinfo{year}{2015}), \eprint{1504.05465}.

\bibitem[{\citenamefont{Troxel and Ishak}(2014)}]{Troxel:2014dba}
\bibinfo{author}{\bibfnamefont{M.~A.} \bibnamefont{Troxel}} \bibnamefont{and} \bibinfo{author}{\bibfnamefont{M.}~\bibnamefont{Ishak}}, \bibinfo{journal}{Phys. Rept.} \textbf{\bibinfo{volume}{558}}, \bibinfo{pages}{1} (\bibinfo{year}{2014}), \eprint{1407.6990}.

\bibitem[{\citenamefont{Lamman et~al.}(2023)\citenamefont{Lamman, Tsaprazi, Shi, \v{S}ar\v{c}evi\'c, Pyne, Legnani, and Ferreira}}]{Lamman:2023hsj}
\bibinfo{author}{\bibfnamefont{C.}~\bibnamefont{Lamman}}, \bibinfo{author}{\bibfnamefont{E.}~\bibnamefont{Tsaprazi}}, \bibinfo{author}{\bibfnamefont{J.}~\bibnamefont{Shi}}, \bibinfo{author}{\bibfnamefont{N.~N.} \bibnamefont{\v{S}ar\v{c}evi\'c}}, \bibinfo{author}{\bibfnamefont{S.}~\bibnamefont{Pyne}}, \bibinfo{author}{\bibfnamefont{E.}~\bibnamefont{Legnani}}, \bibnamefont{and} \bibinfo{author}{\bibfnamefont{T.}~\bibnamefont{Ferreira}} (\bibinfo{year}{2023}), \eprint{2309.08605}.

\bibitem[{\citenamefont{Amendola et~al.}(2018)}]{Amendola:2016saw}
\bibinfo{author}{\bibfnamefont{L.}~\bibnamefont{Amendola}} \bibnamefont{et~al.}, \bibinfo{journal}{Living Reviews in Relativity} \textbf{\bibinfo{volume}{21}}, \bibinfo{pages}{345} (\bibinfo{year}{2018}), \eprint{1606.00180}.

\bibitem[{\citenamefont{Sch\"afer and Merkel}(2017)}]{Schafer:2015qfx}
\bibinfo{author}{\bibfnamefont{B.~M.} \bibnamefont{Sch\"afer}} \bibnamefont{and} \bibinfo{author}{\bibfnamefont{P.~M.} \bibnamefont{Merkel}}, \bibinfo{journal}{Mon. Not. Roy. Astron. Soc.} \textbf{\bibinfo{volume}{470}}, \bibinfo{pages}{3453} (\bibinfo{year}{2017}), \eprint{1506.07366}.

\bibitem[{\citenamefont{Tugendhat and Sch\"afer}(2018)}]{Tugendhat:2017qao}
\bibinfo{author}{\bibfnamefont{T.~M.} \bibnamefont{Tugendhat}} \bibnamefont{and} \bibinfo{author}{\bibfnamefont{B.~M.} \bibnamefont{Sch\"afer}}, \bibinfo{journal}{Mon. Not. Roy. Astron. Soc.} \textbf{\bibinfo{volume}{476}}, \bibinfo{pages}{3460} (\bibinfo{year}{2018}), \eprint{1709.02630}.

\bibitem[{\citenamefont{Reischke and Sch\"afer}(2019)}]{Reischke:2019quo}
\bibinfo{author}{\bibfnamefont{R.}~\bibnamefont{Reischke}} \bibnamefont{and} \bibinfo{author}{\bibfnamefont{B.~M.} \bibnamefont{Sch\"afer}} (\bibinfo{year}{2019}), \eprint{1910.05994}.

\bibitem[{\citenamefont{Zjupa et~al.}(2022)\citenamefont{Zjupa, Sch\"afer, and Hahn}}]{Zjupa:2020kcg}
\bibinfo{author}{\bibfnamefont{J.}~\bibnamefont{Zjupa}}, \bibinfo{author}{\bibfnamefont{B.~M.} \bibnamefont{Sch\"afer}}, \bibnamefont{and} \bibinfo{author}{\bibfnamefont{O.}~\bibnamefont{Hahn}}, \bibinfo{journal}{Mon. Not. Roy. Astron. Soc.} \textbf{\bibinfo{volume}{514}}, \bibinfo{pages}{2049} (\bibinfo{year}{2022}), \eprint{2010.07951}.

\bibitem[{\citenamefont{Delgado et~al.}(2023)}]{Delgado:2023kwx}
\bibinfo{author}{\bibfnamefont{A.~M.} \bibnamefont{Delgado}} \bibnamefont{et~al.}, \bibinfo{journal}{Mon. Not. Roy. Astron. Soc.} \textbf{\bibinfo{volume}{523}}, \bibinfo{pages}{5899} (\bibinfo{year}{2023}), \eprint{2304.12346}.

\bibitem[{\citenamefont{Nelson et~al.}(2018)}]{Nelson:2018uso}
\bibinfo{author}{\bibfnamefont{D.}~\bibnamefont{Nelson}} \bibnamefont{et~al.} (\bibinfo{year}{2018}), \eprint{1812.05609}.

\bibitem[{\citenamefont{{Hildebrandt} et~al.}(2017)\citenamefont{{Hildebrandt}, {Viola}, {Heymans}, {Joudaki}, {Kuijken}, {Blake}, {Erben}, {Joachimi}, {Klaes}, {Miller} et~al.}}]{KiDS450-2017}
\bibinfo{author}{\bibfnamefont{H.}~\bibnamefont{{Hildebrandt}}}, \bibinfo{author}{\bibfnamefont{M.}~\bibnamefont{{Viola}}}, \bibinfo{author}{\bibfnamefont{C.}~\bibnamefont{{Heymans}}}, \bibinfo{author}{\bibfnamefont{S.}~\bibnamefont{{Joudaki}}}, \bibinfo{author}{\bibfnamefont{K.}~\bibnamefont{{Kuijken}}}, \bibinfo{author}{\bibfnamefont{C.}~\bibnamefont{{Blake}}}, \bibinfo{author}{\bibfnamefont{T.}~\bibnamefont{{Erben}}}, \bibinfo{author}{\bibfnamefont{B.}~\bibnamefont{{Joachimi}}}, \bibinfo{author}{\bibfnamefont{D.}~\bibnamefont{{Klaes}}}, \bibinfo{author}{\bibfnamefont{L.}~\bibnamefont{{Miller}}}, \bibnamefont{et~al.}, \bibinfo{journal}{MNRAS} \textbf{\bibinfo{volume}{465}}, \bibinfo{pages}{1454} (\bibinfo{year}{2017}), \eprint{1606.05338}.

\bibitem[{\citenamefont{Johnston et~al.}(2019)}]{Johnston:2018nfi}
\bibinfo{author}{\bibfnamefont{H.}~\bibnamefont{Johnston}} \bibnamefont{et~al.}, \bibinfo{journal}{Astron. Astrophys.} \textbf{\bibinfo{volume}{624}}, \bibinfo{pages}{A30} (\bibinfo{year}{2019}), \eprint{1811.09598}.

\bibitem[{\citenamefont{Samuroff et~al.}(2023)}]{DES:2022vuu}
\bibinfo{author}{\bibfnamefont{S.}~\bibnamefont{Samuroff}} \bibnamefont{et~al.} (\bibinfo{collaboration}{DES}), \bibinfo{journal}{Mon. Not. Roy. Astron. Soc.} \textbf{\bibinfo{volume}{524}}, \bibinfo{pages}{2195} (\bibinfo{year}{2023}), \eprint{2212.11319}.

\bibitem[{\citenamefont{Schaye et~al.}(2015)}]{Schaye:2014tpa}
\bibinfo{author}{\bibfnamefont{J.}~\bibnamefont{Schaye}} \bibnamefont{et~al.}, \bibinfo{journal}{Mon. Not. Roy. Astron. Soc.} \textbf{\bibinfo{volume}{446}}, \bibinfo{pages}{521} (\bibinfo{year}{2015}), \eprint{1407.7040}.

\bibitem[{\citenamefont{Crain et~al.}(2015)}]{Crain:2015poa}
\bibinfo{author}{\bibfnamefont{R.~A.} \bibnamefont{Crain}} \bibnamefont{et~al.}, \bibinfo{journal}{Mon. Not. Roy. Astron. Soc.} \textbf{\bibinfo{volume}{450}}, \bibinfo{pages}{1937} (\bibinfo{year}{2015}), \eprint{1501.01311}.

\bibitem[{\citenamefont{Velliscig et~al.}(2015)}]{Velliscig:2015ixa}
\bibinfo{author}{\bibfnamefont{M.}~\bibnamefont{Velliscig}} \bibnamefont{et~al.}, \bibinfo{journal}{Mon. Not. Roy. Astron. Soc.} \textbf{\bibinfo{volume}{454}}, \bibinfo{pages}{3328} (\bibinfo{year}{2015}), \eprint{1507.06996}.

\bibitem[{\citenamefont{Vogelsberger et~al.}(2014{\natexlab{a}})\citenamefont{Vogelsberger, Genel, Springel, Torrey, Sijacki, Xu, Snyder, Nelson, and Hernquist}}]{Vogelsberger:2014dza}
\bibinfo{author}{\bibfnamefont{M.}~\bibnamefont{Vogelsberger}}, \bibinfo{author}{\bibfnamefont{S.}~\bibnamefont{Genel}}, \bibinfo{author}{\bibfnamefont{V.}~\bibnamefont{Springel}}, \bibinfo{author}{\bibfnamefont{P.}~\bibnamefont{Torrey}}, \bibinfo{author}{\bibfnamefont{D.}~\bibnamefont{Sijacki}}, \bibinfo{author}{\bibfnamefont{D.}~\bibnamefont{Xu}}, \bibinfo{author}{\bibfnamefont{G.~F.} \bibnamefont{Snyder}}, \bibinfo{author}{\bibfnamefont{D.}~\bibnamefont{Nelson}}, \bibnamefont{and} \bibinfo{author}{\bibfnamefont{L.}~\bibnamefont{Hernquist}}, \bibinfo{journal}{Mon. Not. Roy. Astron. Soc.} \textbf{\bibinfo{volume}{444}}, \bibinfo{pages}{1518} (\bibinfo{year}{2014}{\natexlab{a}}), \eprint{1405.2921}.

\bibitem[{\citenamefont{Vogelsberger et~al.}(2014{\natexlab{b}})\citenamefont{Vogelsberger, Genel, Springel, Torrey, Sijacki, Xu, Snyder, Bird, Nelson, and Hernquist}}]{Vogelsberger:2014kha}
\bibinfo{author}{\bibfnamefont{M.}~\bibnamefont{Vogelsberger}}, \bibinfo{author}{\bibfnamefont{S.}~\bibnamefont{Genel}}, \bibinfo{author}{\bibfnamefont{V.}~\bibnamefont{Springel}}, \bibinfo{author}{\bibfnamefont{P.}~\bibnamefont{Torrey}}, \bibinfo{author}{\bibfnamefont{D.}~\bibnamefont{Sijacki}}, \bibinfo{author}{\bibfnamefont{D.}~\bibnamefont{Xu}}, \bibinfo{author}{\bibfnamefont{G.~F.} \bibnamefont{Snyder}}, \bibinfo{author}{\bibfnamefont{S.}~\bibnamefont{Bird}}, \bibinfo{author}{\bibfnamefont{D.}~\bibnamefont{Nelson}}, \bibnamefont{and} \bibinfo{author}{\bibfnamefont{L.}~\bibnamefont{Hernquist}}, \bibinfo{journal}{Nature} \textbf{\bibinfo{volume}{509}}, \bibinfo{pages}{177} (\bibinfo{year}{2014}{\natexlab{b}}), \eprint{1405.1418}.

\bibitem[{\citenamefont{Hilbert et~al.}(2017)\citenamefont{Hilbert, Xu, Schneider, Springel, Vogelsberger, and Hernquist}}]{Hilbert:2016ylf}
\bibinfo{author}{\bibfnamefont{S.}~\bibnamefont{Hilbert}}, \bibinfo{author}{\bibfnamefont{D.}~\bibnamefont{Xu}}, \bibinfo{author}{\bibfnamefont{P.}~\bibnamefont{Schneider}}, \bibinfo{author}{\bibfnamefont{V.}~\bibnamefont{Springel}}, \bibinfo{author}{\bibfnamefont{M.}~\bibnamefont{Vogelsberger}}, \bibnamefont{and} \bibinfo{author}{\bibfnamefont{L.}~\bibnamefont{Hernquist}}, \bibinfo{journal}{Mon. Not. Roy. Astron. Soc.} \textbf{\bibinfo{volume}{468}}, \bibinfo{pages}{790} (\bibinfo{year}{2017}), \eprint{1606.03216}.

\bibitem[{\citenamefont{Tenneti et~al.}(2015)\citenamefont{Tenneti, Singh, Mandelbaum, Di~Matteo, Feng, and Khandai}}]{Tenneti:2014bca}
\bibinfo{author}{\bibfnamefont{A.}~\bibnamefont{Tenneti}}, \bibinfo{author}{\bibfnamefont{S.}~\bibnamefont{Singh}}, \bibinfo{author}{\bibfnamefont{R.}~\bibnamefont{Mandelbaum}}, \bibinfo{author}{\bibfnamefont{T.}~\bibnamefont{Di~Matteo}}, \bibinfo{author}{\bibfnamefont{Y.}~\bibnamefont{Feng}}, \bibnamefont{and} \bibinfo{author}{\bibfnamefont{N.}~\bibnamefont{Khandai}}, \bibinfo{journal}{Mon. Not. Roy. Astron. Soc.} \textbf{\bibinfo{volume}{448}}, \bibinfo{pages}{3522} (\bibinfo{year}{2015}), \eprint{1409.7297}.

\bibitem[{\citenamefont{Dav\'e et~al.}(2019)\citenamefont{Dav\'e, Angl\'es-Alc\'azar, Narayanan, Li, Rafieferantsoa, and Appleby}}]{Dave:2019yyq}
\bibinfo{author}{\bibfnamefont{R.}~\bibnamefont{Dav\'e}}, \bibinfo{author}{\bibfnamefont{D.}~\bibnamefont{Angl\'es-Alc\'azar}}, \bibinfo{author}{\bibfnamefont{D.}~\bibnamefont{Narayanan}}, \bibinfo{author}{\bibfnamefont{Q.}~\bibnamefont{Li}}, \bibinfo{author}{\bibfnamefont{M.~H.} \bibnamefont{Rafieferantsoa}}, \bibnamefont{and} \bibinfo{author}{\bibfnamefont{S.}~\bibnamefont{Appleby}}, \bibinfo{journal}{Mon. Not. Roy. Astron. Soc.} \textbf{\bibinfo{volume}{486}}, \bibinfo{pages}{2827} (\bibinfo{year}{2019}), \eprint{1901.10203}.

\bibitem[{\citenamefont{Blazek et~al.}(2019)\citenamefont{Blazek, MacCrann, Troxel, and Fang}}]{Blazek:2017wbz}
\bibinfo{author}{\bibfnamefont{J.}~\bibnamefont{Blazek}}, \bibinfo{author}{\bibfnamefont{N.}~\bibnamefont{MacCrann}}, \bibinfo{author}{\bibfnamefont{M.~A.} \bibnamefont{Troxel}}, \bibnamefont{and} \bibinfo{author}{\bibfnamefont{X.}~\bibnamefont{Fang}}, \bibinfo{journal}{Phys. Rev. D} \textbf{\bibinfo{volume}{100}}, \bibinfo{pages}{103506} (\bibinfo{year}{2019}), \eprint{1708.09247}.

\bibitem[{\citenamefont{Vlah et~al.}(2020)\citenamefont{Vlah, Chisari, and Schmidt}}]{Vlah:2019byq}
\bibinfo{author}{\bibfnamefont{Z.}~\bibnamefont{Vlah}}, \bibinfo{author}{\bibfnamefont{N.~E.} \bibnamefont{Chisari}}, \bibnamefont{and} \bibinfo{author}{\bibfnamefont{F.}~\bibnamefont{Schmidt}}, \bibinfo{journal}{JCAP} \textbf{\bibinfo{volume}{01}}, \bibinfo{pages}{025} (\bibinfo{year}{2020}), \eprint{1910.08085}.

\bibitem[{\citenamefont{Chen and Kokron}(2024)}]{Chen:2023yyb}
\bibinfo{author}{\bibfnamefont{S.-F.} \bibnamefont{Chen}} \bibnamefont{and} \bibinfo{author}{\bibfnamefont{N.}~\bibnamefont{Kokron}}, \bibinfo{journal}{JCAP} \textbf{\bibinfo{volume}{01}}, \bibinfo{pages}{027} (\bibinfo{year}{2024}), \eprint{2309.16761}.

\bibitem[{\citenamefont{Ghosh et~al.}(2021)\citenamefont{Ghosh, Durrer, and Schaefer}}]{Ghosh:2020zfa}
\bibinfo{author}{\bibfnamefont{B.}~\bibnamefont{Ghosh}}, \bibinfo{author}{\bibfnamefont{R.}~\bibnamefont{Durrer}}, \bibnamefont{and} \bibinfo{author}{\bibfnamefont{B.~M.} \bibnamefont{Schaefer}}, \bibinfo{journal}{Mon. Not. Roy. Astron. Soc.} \textbf{\bibinfo{volume}{505}}, \bibinfo{pages}{2594} (\bibinfo{year}{2021}), \eprint{2005.04604}.

\bibitem[{\citenamefont{Navarro et~al.}(1996)\citenamefont{Navarro, Frenk, and White}}]{Navarro:1995iw}
\bibinfo{author}{\bibfnamefont{J.~F.} \bibnamefont{Navarro}}, \bibinfo{author}{\bibfnamefont{C.~S.} \bibnamefont{Frenk}}, \bibnamefont{and} \bibinfo{author}{\bibfnamefont{S.~D.~M.} \bibnamefont{White}}, \bibinfo{journal}{Astrophys. J.} \textbf{\bibinfo{volume}{462}}, \bibinfo{pages}{563} (\bibinfo{year}{1996}), \eprint{astro-ph/9508025}.

\bibitem[{\citenamefont{Jagvaral et~al.}(2022)\citenamefont{Jagvaral, Singh, and Mandelbaum}}]{Jagvaral:2022zto}
\bibinfo{author}{\bibfnamefont{Y.}~\bibnamefont{Jagvaral}}, \bibinfo{author}{\bibfnamefont{S.}~\bibnamefont{Singh}}, \bibnamefont{and} \bibinfo{author}{\bibfnamefont{R.}~\bibnamefont{Mandelbaum}}, \bibinfo{journal}{Mon. Not. Roy. Astron. Soc.} \textbf{\bibinfo{volume}{514}}, \bibinfo{pages}{1021} (\bibinfo{year}{2022}), \eprint{2202.08849}.

\bibitem[{\citenamefont{Heymans et~al.}(2013)}]{Heymans:2013fya}
\bibinfo{author}{\bibfnamefont{C.}~\bibnamefont{Heymans}} \bibnamefont{et~al.}, \bibinfo{journal}{Mon. Not. Roy. Astron. Soc.} \textbf{\bibinfo{volume}{432}}, \bibinfo{pages}{2433} (\bibinfo{year}{2013}), \eprint{1303.1808}.

\bibitem[{\citenamefont{Kilbinger et~al.}(2009)}]{Kilbinger:2008gk}
\bibinfo{author}{\bibfnamefont{M.}~\bibnamefont{Kilbinger}} \bibnamefont{et~al.}, \bibinfo{journal}{Astron. Astrophys.} \textbf{\bibinfo{volume}{497}}, \bibinfo{pages}{677} (\bibinfo{year}{2009}), \eprint{0810.5129}.

\bibitem[{\citenamefont{Blazek et~al.}(2011)\citenamefont{Blazek, McQuinn, and Seljak}}]{Blazek:2011xq}
\bibinfo{author}{\bibfnamefont{J.}~\bibnamefont{Blazek}}, \bibinfo{author}{\bibfnamefont{M.}~\bibnamefont{McQuinn}}, \bibnamefont{and} \bibinfo{author}{\bibfnamefont{U.}~\bibnamefont{Seljak}}, \bibinfo{journal}{JCAP} \textbf{\bibinfo{volume}{05}}, \bibinfo{pages}{010} (\bibinfo{year}{2011}), \eprint{1101.4017}.

\bibitem[{\citenamefont{Blazek et~al.}(2015)\citenamefont{Blazek, Vlah, and Seljak}}]{Blazek:2015lfa}
\bibinfo{author}{\bibfnamefont{J.}~\bibnamefont{Blazek}}, \bibinfo{author}{\bibfnamefont{Z.}~\bibnamefont{Vlah}}, \bibnamefont{and} \bibinfo{author}{\bibfnamefont{U.}~\bibnamefont{Seljak}}, \bibinfo{journal}{JCAP} \textbf{\bibinfo{volume}{08}}, \bibinfo{pages}{015} (\bibinfo{year}{2015}), \eprint{1504.02510}.

\bibitem[{\citenamefont{Giesel et~al.}(2022{\natexlab{a}})\citenamefont{Giesel, Ghosh, and Schaefer}}]{Giesel:2021ikc}
\bibinfo{author}{\bibfnamefont{E.~S.} \bibnamefont{Giesel}}, \bibinfo{author}{\bibfnamefont{B.}~\bibnamefont{Ghosh}}, \bibnamefont{and} \bibinfo{author}{\bibfnamefont{B.~M.} \bibnamefont{Schaefer}}, \bibinfo{journal}{Mon. Not. Roy. Astron. Soc.} \textbf{\bibinfo{volume}{510}}, \bibinfo{pages}{2773} (\bibinfo{year}{2022}{\natexlab{a}}), \eprint{2107.09000}.

\bibitem[{\citenamefont{Camelio and Lombardi}(2015)}]{Camelio:2015gda}
\bibinfo{author}{\bibfnamefont{G.}~\bibnamefont{Camelio}} \bibnamefont{and} \bibinfo{author}{\bibfnamefont{M.}~\bibnamefont{Lombardi}}, \bibinfo{journal}{Astron. Astrophys.} \textbf{\bibinfo{volume}{575}}, \bibinfo{pages}{A113} (\bibinfo{year}{2015}), \eprint{1501.03014}.

\bibitem[{\citenamefont{Child et~al.}(2018)\citenamefont{Child, Habib, Heitmann, Frontiere, Finkel, Pope, and Morozov}}]{Child:2018skq}
\bibinfo{author}{\bibfnamefont{H.~L.} \bibnamefont{Child}}, \bibinfo{author}{\bibfnamefont{S.}~\bibnamefont{Habib}}, \bibinfo{author}{\bibfnamefont{K.}~\bibnamefont{Heitmann}}, \bibinfo{author}{\bibfnamefont{N.}~\bibnamefont{Frontiere}}, \bibinfo{author}{\bibfnamefont{H.}~\bibnamefont{Finkel}}, \bibinfo{author}{\bibfnamefont{A.}~\bibnamefont{Pope}}, \bibnamefont{and} \bibinfo{author}{\bibfnamefont{V.}~\bibnamefont{Morozov}}, \bibinfo{journal}{Astrophys. J.} \textbf{\bibinfo{volume}{859}}, \bibinfo{pages}{55} (\bibinfo{year}{2018}), \eprint{1804.10199}.

\bibitem[{\citenamefont{{Einasto}}(1965)}]{1965TrAlm...5...87E}
\bibinfo{author}{\bibfnamefont{J.}~\bibnamefont{{Einasto}}}, \bibinfo{journal}{Trudy Astrofizicheskogo Instituta Alma-Ata} \textbf{\bibinfo{volume}{5}}, \bibinfo{pages}{87} (\bibinfo{year}{1965}).

\bibitem[{\citenamefont{Giesel et~al.}(2022{\natexlab{b}})\citenamefont{Giesel, Ghosh, and Sch\"afer}}]{Giesel:2022org}
\bibinfo{author}{\bibfnamefont{E.~S.} \bibnamefont{Giesel}}, \bibinfo{author}{\bibfnamefont{B.}~\bibnamefont{Ghosh}}, \bibnamefont{and} \bibinfo{author}{\bibfnamefont{B.~M.} \bibnamefont{Sch\"afer}}, \bibinfo{journal}{Mon. Not. Roy. Astron. Soc.} \textbf{\bibinfo{volume}{518}}, \bibinfo{pages}{5490} (\bibinfo{year}{2022}{\natexlab{b}}), \eprint{2208.07197}.

\bibitem[{\citenamefont{Dehnen et~al.}(2006)\citenamefont{Dehnen, McLaughlin, and Sachania}}]{Dehnen:2006cm}
\bibinfo{author}{\bibfnamefont{W.}~\bibnamefont{Dehnen}}, \bibinfo{author}{\bibfnamefont{D.}~\bibnamefont{McLaughlin}}, \bibnamefont{and} \bibinfo{author}{\bibfnamefont{J.}~\bibnamefont{Sachania}}, \bibinfo{journal}{Mon. Not. Roy. Astron. Soc.} \textbf{\bibinfo{volume}{369}}, \bibinfo{pages}{1688} (\bibinfo{year}{2006}), \eprint{astro-ph/0603825}.

\bibitem[{\citenamefont{Lin and Li}(2019)}]{Lin:2019yux}
\bibinfo{author}{\bibfnamefont{H.-N.} \bibnamefont{Lin}} \bibnamefont{and} \bibinfo{author}{\bibfnamefont{X.}~\bibnamefont{Li}}, \bibinfo{journal}{Mon. Not. Roy. Astron. Soc.} \textbf{\bibinfo{volume}{487}}, \bibinfo{pages}{5679} (\bibinfo{year}{2019}), \eprint{1906.08419}.

\bibitem[{\citenamefont{Goodwin et~al.}(1997)\citenamefont{Goodwin, Gribbin, and Hendry}}]{Goodwin:1997ys}
\bibinfo{author}{\bibfnamefont{S.~P.} \bibnamefont{Goodwin}}, \bibinfo{author}{\bibfnamefont{J.}~\bibnamefont{Gribbin}}, \bibnamefont{and} \bibinfo{author}{\bibfnamefont{M.~A.} \bibnamefont{Hendry}} (\bibinfo{year}{1997}), \eprint{astro-ph/9704216}.

\bibitem[{\citenamefont{{Goodwin} et~al.}(1998)\citenamefont{{Goodwin}, {Gribbin}, and {Hendry}}}]{1998Obs...118..201G}
\bibinfo{author}{\bibfnamefont{S.~P.} \bibnamefont{{Goodwin}}}, \bibinfo{author}{\bibfnamefont{J.}~\bibnamefont{{Gribbin}}}, \bibnamefont{and} \bibinfo{author}{\bibfnamefont{M.~A.} \bibnamefont{{Hendry}}}, \bibinfo{journal}{The Observatory} \textbf{\bibinfo{volume}{118}}, \bibinfo{pages}{201} (\bibinfo{year}{1998}).

\bibitem[{\citenamefont{Brown et~al.}(2002)\citenamefont{Brown, Taylor, Hambly, and Dye}}]{Brown:2000gt}
\bibinfo{author}{\bibfnamefont{M.~L.} \bibnamefont{Brown}}, \bibinfo{author}{\bibfnamefont{A.~N.} \bibnamefont{Taylor}}, \bibinfo{author}{\bibfnamefont{N.~C.} \bibnamefont{Hambly}}, \bibnamefont{and} \bibinfo{author}{\bibfnamefont{S.}~\bibnamefont{Dye}}, \bibinfo{journal}{Mon. Not. Roy. Astron. Soc.} \textbf{\bibinfo{volume}{333}}, \bibinfo{pages}{501} (\bibinfo{year}{2002}), \eprint{astro-ph/0009499}.

\bibitem[{\citenamefont{Bretonni\`ere et~al.}(2023)}]{Euclid:2022vtv}
\bibinfo{author}{\bibfnamefont{H.}~\bibnamefont{Bretonni\`ere}} \bibnamefont{et~al.} (\bibinfo{collaboration}{Euclid}), \bibinfo{journal}{Astron. Astrophys.} \textbf{\bibinfo{volume}{671}}, \bibinfo{pages}{A102} (\bibinfo{year}{2023}), \eprint{2209.12907}.

\bibitem[{\citenamefont{Valenti et~al.}(2018)\citenamefont{Valenti, Zoccali, Mucciarelli, Gonzalez, Surot, Minniti, Rejkuba, Pasquini, Fiorentino, Bono et~al.}}]{Valenti_2018}
\bibinfo{author}{\bibfnamefont{E.}~\bibnamefont{Valenti}}, \bibinfo{author}{\bibfnamefont{M.}~\bibnamefont{Zoccali}}, \bibinfo{author}{\bibfnamefont{A.}~\bibnamefont{Mucciarelli}}, \bibinfo{author}{\bibfnamefont{O.~A.} \bibnamefont{Gonzalez}}, \bibinfo{author}{\bibfnamefont{F.}~\bibnamefont{Surot}}, \bibinfo{author}{\bibfnamefont{D.}~\bibnamefont{Minniti}}, \bibinfo{author}{\bibfnamefont{M.}~\bibnamefont{Rejkuba}}, \bibinfo{author}{\bibfnamefont{L.}~\bibnamefont{Pasquini}}, \bibinfo{author}{\bibfnamefont{G.}~\bibnamefont{Fiorentino}}, \bibinfo{author}{\bibfnamefont{G.}~\bibnamefont{Bono}}, \bibnamefont{et~al.}, \bibinfo{journal}{Astronomy \& Astrophysics} \textbf{\bibinfo{volume}{616}}, \bibinfo{pages}{A83} (\bibinfo{year}{2018}), ISSN \bibinfo{issn}{1432-0746}.

\bibitem[{\citenamefont{Bartelmann and Schneider}(2001)}]{bartelmann_weak_2001}
\bibinfo{author}{\bibfnamefont{M.}~\bibnamefont{Bartelmann}} \bibnamefont{and} \bibinfo{author}{\bibfnamefont{P.}~\bibnamefont{Schneider}}, \bibinfo{journal}{Physics Reports} \textbf{\bibinfo{volume}{340}}, \bibinfo{pages}{291} (\bibinfo{year}{2001}).

\bibitem[{\citenamefont{Ciarlariello et~al.}(2015)\citenamefont{Ciarlariello, Crittenden, and Pace}}]{Ciarlariello:2014qva}
\bibinfo{author}{\bibfnamefont{S.}~\bibnamefont{Ciarlariello}}, \bibinfo{author}{\bibfnamefont{R.}~\bibnamefont{Crittenden}}, \bibnamefont{and} \bibinfo{author}{\bibfnamefont{F.}~\bibnamefont{Pace}}, \bibinfo{journal}{Mon. Not. Roy. Astron. Soc.} \textbf{\bibinfo{volume}{449}}, \bibinfo{pages}{2059} (\bibinfo{year}{2015}), \eprint{1412.4606}.

\bibitem[{\citenamefont{Ciarlariello and Crittenden}(2016)}]{Ciarlariello:2016hxr}
\bibinfo{author}{\bibfnamefont{S.}~\bibnamefont{Ciarlariello}} \bibnamefont{and} \bibinfo{author}{\bibfnamefont{R.}~\bibnamefont{Crittenden}}, \bibinfo{journal}{Mon. Not. Roy. Astron. Soc.} \textbf{\bibinfo{volume}{463}}, \bibinfo{pages}{740} (\bibinfo{year}{2016}), \eprint{1607.08784}.

\bibitem[{\citenamefont{Johnston et~al.}(2023)\citenamefont{Johnston, Westbeek, Weide, Chisari, Dubois, Devriendt, and Pichon}}]{Johnston:2022nbv}
\bibinfo{author}{\bibfnamefont{H.}~\bibnamefont{Johnston}}, \bibinfo{author}{\bibfnamefont{D.~S.} \bibnamefont{Westbeek}}, \bibinfo{author}{\bibfnamefont{S.}~\bibnamefont{Weide}}, \bibinfo{author}{\bibfnamefont{N.~E.} \bibnamefont{Chisari}}, \bibinfo{author}{\bibfnamefont{Y.}~\bibnamefont{Dubois}}, \bibinfo{author}{\bibfnamefont{J.}~\bibnamefont{Devriendt}}, \bibnamefont{and} \bibinfo{author}{\bibfnamefont{C.}~\bibnamefont{Pichon}}, \bibinfo{journal}{Mon. Not. Roy. Astron. Soc.} \textbf{\bibinfo{volume}{520}}, \bibinfo{pages}{1541} (\bibinfo{year}{2023}), \eprint{2209.11063}.

\end{thebibliography}

\end{document}